\documentclass[st,onecolumn]{jpsj3}
\usepackage{txfonts}
\usepackage{graphicx}
\usepackage{amssymb, amsfonts, amsmath}
\usepackage{dcolumn}
\usepackage{bm}
\usepackage{color}
\usepackage[normalem]{ulem}

\title{First-principles theory of magnetic multipoles
in condensed matter systems}

\author{Michi-To Suzuki$^1$\thanks{michito.suzuki@riken.jp}, Hiroaki
Ikeda$^2$, and Peter M.\ Oppeneer$^3$}
\inst{$^1$RIKEN Center for Emergent Matter Science, 2-1 Hirosawa, Wako, Saitama 351-0198, Japan \\
$^2$Department of Physics, Ritsumeikan University, Kusatsu, Shiga 525-8577, Japan \\
$^3$Department of Physics and Astronomy, Uppsala University, P.\,O.\ Box
516, SE-751 20 Uppsala, Sweden} 

\abst{
  The multipole concept, which characterizes the spacial distribution
 of scalar and vector objects by their angular dependence, has already become widely used in various areas of physics.
  In recent years it has become employed to systematically classify the
  anisotropic distribution of electrons and magnetization around atoms 
  in solid state materials.  
  This has been fuelled by the discovery of several physical phenomena
  that exhibit unusual higher rank multipole moments, beyond that of the
  conventional  degrees of freedom as charge and magnetic dipole moment.
  Moreover, the higher rank electric/magnetic multipole moments have been suggested as promising
  order parameters in exotic hidden order phases.  
  While the experimental investigations of such anomalous phases
  have provided encouraging observations of multipolar order,
  theoretical approaches have developed at a slower pace.
  In particular, a materials' specific theory has been missing. 
 The multipole concept has furthermore been recognized as the key quantity which
 characterizes the resultant configuration of magnetic moments in a cluster of atomic moments.
  This cluster multipole moment has then been
  introduced as macroscopic order parameter for a noncollinear
  antiferromagnetic structure in crystals that can explain unusual
 physical phenomena whose appearance is determined by the
 magnetic point group symmetry.
  It is the purpose of this review to discuss the recent developments in
  the first-principles theory investigating multipolar degrees of freedom
  in condensed matter systems.
  These recent developments exemplify that \textit{ab initio} electronic
  structure calculations can unveil detailed insight in the mechanism of
  physical phenomena caused by the unconventional, multipole degree of freedom.

}

\begin{document}
\maketitle

\section{Introduction}
The multipole formulation has proven to be a very fruitful as well as foundational concept in physics \cite{raab2004}. 
More than a century ago the single electron was discovered to have a unique monopole charge. Diatomic molecules became classified after their electric dipole moment. 
The possible existence of the Dirac magnetic monopole would have
fundamental consequences  \cite{dirac1931} and has been much
discussed. On the basis of the Maxwell equations the dynamics of
electrically charged bodies became generalized, leading to the multipole
description of their charge distributions \cite{jackson}. The multipole
description has subsequently become used in many branches of physics,
including nuclear physics \cite{blatt1952_book,blin1956} and solid state
physics (see, e.g., Refs.\ \cite{kuramoto2008,kusunose2008,kuramoto2009,santini2009}). 
 In this review we focus on the appearance of electric or magnetic multipole moments in condensed matter systems.
%

Condensed matter systems display a richness of physical phenomena, such as e.g.\ strongly correlated electron behavior, complex magnetic orders, orbital order, unusual forms of superconductivity and quantum critical behavior. This diversity of phenomena  can be attributed to the multiple degrees of freedom available to the electronic system that, in itself or in conjunction to lattice degrees of freedom, can have access to entangled spin and orbital states. 
 In particular, the electron clouds around nuclei in crystalline materials
can be systematically characterized by using the multipole expansion. Since electrons on
filled atomic shells lead only to spherical charge distributions, the
electrons in open shells are responsible for the anisotropic charge and
spin distributions and thus for the eventual appearance of multipolar
order. It turns out, however, that the appearance of true multipolar order, that is, of rank higher than dipolar order, is actually rare in
solid state compounds. Thus far, true multipolar order has
been discovered, or predicted, for about a dozen crystalline materials,
such as, 
CeB$_6$~\cite{erkelens1987, shiina1997, nakao2001, zaharko2003, tanaka2005, matsumura2009, akbari2012, matsumura2012, friemel2012, matsumura2013, jang2014, cameron2016},
$R_3$Pd$_{20}$$X_6$ ($R$: Ce, Pr, $X$: Si, Ge)~\cite{kitagawa1996, kitagawa1998, kitagawa2000, goto2002, nemoto2002, nemoto2003, goto2009, deen2010, ano2012, steinke2013, portnichenko2016},
PrPb$_3$~\cite{bucher1974, tayama2001, onimaru2005, onimaru2006, ito2007, sakai2008, onishi2008, ito2009, kawae2013, kubo2017}, 
PrMg$_3$, \cite{tanida2006, araki2011, morie2009, araki2012}
PrFe$_4$P$_{12}$~\cite{iwasa2002, sugawara2002, hao2003, aoki2005, hao2005, hotta2005, sato2007, sakai2007, kikuchi2007, ishii2008, sugawara2008, sato2009, tayama2009, iwasa2012, hotta2012},
PrRu$_4$P$_{12}$~\cite{iwasa2005, takimoto2006a, takimoto2006b, harima2008, sakakibara2008, shiina2010, shiina2011, okamura2012, goho2015},
PrOs$_4$Sb$_{12}$~\cite{aoki2002, tayama2003, kohgi2003, shiina2004, aoki2005, yogi2006, iwasa2009},
SmRu$_4$P$_{12}$~\cite{kiss2006, masaki2006, aoki2007, shiina2013, matsumura2014, matsumura2016}, 
Pr$T_2$$X_{20}$ ($T$: Ir, Rh, $X$: Zn; $T$: V, Ti, $X$:
Al)~\cite{onimaru2011, ishii2011, sakai2012, onimaru2012,
matsubayashi2012, ishii2013, tsujimoto2014, tsuruta2015, kubo2015,
matsumoto2015, shimura2015, taniguchi2016, onimaru2016, onimaru2016a, iwasa2017}, 
DyB$_2$C$_2$~\cite{yamauchi1999, tanaka1999, hirota2000, matsumura2002, tanaka2004, lovesey2005, staub2005, matsumura2005,  mulders2006, fernandez2010a, princep2011}, 
HoB$_2$C$_2$~\cite{yanagisawa2005a, yanagisawa2005b, hillier2007, yamauchi2012}
DyPd$_3$S$_4$~\cite{matsuoka2001, matsuoka2002, keller2004, matsuoka2007}, 
CePd$_3$S$_4$~\cite{abe1999, matsuoka2008, michimura2012}
UO$_2$~\cite{santini2009, burlet1986, wilkins2006, matteo2007, carretta2010, caciuffo2011, mtsuzuki2013}, 
NpO$_2$~\cite{santini2009, erdos1980, caciuffo1987, kopmann1998, santini2000, paixao2002, caciuffo2003, tokunaga2005, kubo2005c, sakai2005, santini2006, tokunaga2006, mtsuzuki2010}, 
AmO$_2$~\cite{hotta2009, hotta2012, tokunaga2014, walstedt2014},
UPd$_3$~\cite{mcewen1998, lingg1999, tokiwa2001, mcmorrow2001,
schenck2002, sikora2006, walker2006, fujimoto2007, walker2008,
fernandez2010b, walker2011}, and
URu$_2$Si$_2$~\cite{santini1994, kiss2005, hanzawa2005, hanzawa2007, cricchio2009, haule2009, harima2010, thalmeier2011, kusunose2011, walker2011a, mydosh2011, ikeda2012, ressouche2012, khalyavin2014, chandra2014, kung2015, kung2016}.


These compounds have in common that the exotic multipolar electric or
magnetic distribution occurs on the lanthanide or actinide ion and are
thus generated by the $f$ electrons. More general, typical features of
$f$ electron systems are localized character, that tends to enforce
the strong electron-electron repulsion, a significant orbital degree
of freedom, allowing for anisotropic orbital states, and strong spin-orbit (S-O) coupling which leads to entangled spin and orbital moments. 

 A major question is which theoretical description is suitable to unveil
insight in the origin of the multipolar order in crystal systems.
Single ion crystal-electric field (CEF) theories could previously
provide understanding of aspects of multipolar ordered ionic states,
often in relation to the local point group symmetry and assumptions made
regarding the CEF level
scheme~\cite{shiina1997,kiss2003,kiss2005,kusunose2008,kuramoto2008,kuramoto2009,santini2009}. 
However, this approach has limitations as it considers basically only
the atomic $f$-orbitals yet without materials' specific aspects, such as
e.g.\ hybridization, which can significantly modify the materials'
properties. 
Conversely, the density-functional theory (DFT) (see, e.g.\ Ref.\ \cite{jones1989}) offers a first-principles framework for the efficient and accurate calculation of materials' specific properties.
However, this framework employing the common approximations used in electronic structure calculations, such as the
local spin density approximation (LSDA) and generalized gradient approximation (GGA), often fails to capture important aspect of the $f$ electronic states due to the incompletely captured strong electron correlation effect in the LSDA or GGA exchange-correlation interaction. 
The weakness to describe sufficiently the strong $f$-electron correlation immediately leads to a difficulty to
 obtain the correct ground state of a strongly correlated electron system, since the strong electron
 correlation is usually inseparable from the complex orbital degree of freedom.
 To treat multipolar degrees of freedom from first principles, it is thus needed to treat strong local electron correlations, spin-orbit interaction, and
 the multi-orbital $f$ electron character on an equal footing.


Recently, first-principles electronic structure calculations have been
shown to be suitable for investigation of the multipole ordered states
in correlated solid state materials~\cite{bultmark2009, cricchio2009, haule2009,
mtsuzuki2010, ikeda2012, mtsuzuki2013, mtsuzuki2014, goho2015}.
Hence, the way to consider multipole systems in condensed matter physics is now shifting more to 
the materials' specific aspect as compared to the single-ion treatment in CEF theory.
%
  Such first-principles calculations could recently provide insight in the
 multipole order phases appearing in the actinide dioxides~\cite{mtsuzuki2010, mtsuzuki2013} and, more generally, demonstrated
 the possibility of performing fully first-principles study of
 complex ordered phase in $f$ electron compounds.  Combining the first-principles 
 calculations with a group theoretical analysis led to a powerful tool 
 to investigate hidden order parameters with the multipole degree of freedom~\cite{mtsuzuki2014}. 
 While these \textit{ab initio} calculations describe the anomalously ordered zero temperature ground state,
 extensions to the description of e.g.\ 
 phase diagrams have been recently undertaken.
  Reliable estimations of physical quantities related to the multipolar
  degree of freedom is a difficult task, but the situation has significantly
  improved by recent developments of a first-principles approach to construct tight-binding models~\cite{mostofi2008}.
  In addition, calculations of multipole fluctuations have provided important
 information about instability for the phase transition to a hidden order phase 
 and unconventional superconductivity of heavy Fermion
 compounds~\cite{ikeda2012, ikeda2015}.
 %
%
Lastly, the development of multipole theory has recently
led to new areas of condensed matter physics where the concept is becoming employed.
 As the multipole classification scheme can be generalized to systems characterized
 by a specific point group symmetry, multipole theory is becoming used
 for the classification of off-diagonal response phenomena, such as the
 electric-magnetic effect and the anomalous Hall effect in
 antiferromagnets~\cite{ederer2007,spaldin2008,spaldin2013,hayami2014,hayami2016,mtsuzuki2017,sumita2017}.

In the following, we discuss first in Sec.\ 2 the basic concept of the multipole degree of freedom.
In Sec.\ \ref{Sec:MPorder}, the first-principles
approach to investigate the multipolar ordered phases is discussed first, after which
we review the first-principles calculation
of multipole ordered phases, mainly focusing on the actinide
dioxides. 
 The relation between the electronic structures obtained from the first-principles calculations and the CEF analysis is provided.
In Sec.\ \ref{Sec:MPfluct}, we discuss the first-principles calculations
of multipole fluctuations and review the study 
of superconducting pairing mediated by multipole fluctuations in
CeCu$_2$Si$_2$.
In Sec.\ \ref{Sec:HiddenOrder}, we focus on the long-standing problem of
the hidden-order phase of URu$_2$Si$_2$ and discuss the first-principles
approach to identify the possible order parameters of this phase, using
first-principles calculations and group theoretical analysis. 
Lastly, in Sec.\ \ref{Sec:ClusterMP} we provide an outlook on the
emerging multipolar theory to describe transport phenomena with the
introduction of the study of the anomalous Hall effect in the
antiferromagnetic phases of Mn$_3$$Z$ ($Z$=Sn, Ge), and conclude 
this survey in Sec.\ 7.

\section{Multipole moments in condensed matter system}
\subsection{General concept of the multipole moment}

 
 The multipole moments are mathematically defined as expansion coefficients of
 the multipole expansion of $f(\theta,\phi)$ defined on the surface space
 ($0\le\theta\le\pi$, $0\le\phi <2\pi$) satisfying the condition $\int_{0}^{2\pi}d\phi\int_{0}^{\pi}d\theta\sin \theta |f(\theta,\phi)|^2<\infty$~\cite{varshalovich1988_book}.
  When the function $f(\theta,\phi)$ is expanded as:
\begin{eqnarray}
 f(\theta,\phi)=\sum_{\ell =0}^{\infty}\sum_{m=-\ell}^{\ell}a_{\ell m}
  Y_{\ell m}(\theta,\phi)\ ,
\label{Eq:GeneralMPexpans}
\end{eqnarray}
  the expansion coefficient $a_{\ell m}$ is the multipole moment.
  The multipole moments are thus obtained from
\begin{eqnarray}
 a_{\ell m}=\int_{0}^{2\pi}d\phi\int_{0}^{\pi}d\theta\,  {\rm sin} \theta \,
  Y^{*}_{\ell m}(\theta,\phi)f(\theta,\phi)\ ,
\label{Eq:GeneralMPmom}
\end{eqnarray}
 where $\ell$ is called the rank of multipole moments, using the
 orthogonality relation of spherical harmonics.
 Since the spherical harmonics form a complete orthonormal basis set on
 the sphere space, the function $f(\theta,\phi)$ is fully identified
 with the multipole moments from Eq.\ (\ref{Eq:GeneralMPexpans}).

 In condensed matter physics, the angular distribution of a
 spin-polarized electron cloud around a nucleus can be characterized by
 spherical harmonics.
 The electric multipole moments, which
measure the anisotropy of charge distribution, are defined as the
projection of the charge density $\rho_e({\bm r})$ onto the spherical
harmonics:
\begin{eqnarray}
 \mathcal{Q}_{\ell m} \equiv  \sqrt{\frac{4\pi}{2\ell +1}}\int d{\bm r} \, \Bigl(r^{\ell}Y_{\ell m}^{*}(\theta,\phi)\Bigr)\,  \rho_e({\bm
  r}) ,
 \label{Eq:ElecMP}
\end{eqnarray}
where the integration is over a suitably chosen atomic volume.
   The magnetic multipole moment is defined by introducing the magnetic
   monopole charge, which is related to the magnetization distribution defined by $\rho_m
   ({\bm r}) = -{\bm \nabla} \cdot {\bm m} ({\bm r})$, and accordingly
   gives 
\begin{eqnarray}
 \mathcal{M}_{\ell m}\! \! & \! \equiv \! & \! \!
  \sqrt{\frac{4\pi}{2\ell +1}}\int d{\bm r\, }\Bigl(r^{\ell} Y_{\ell
  m}^{*}(\theta,\phi)\Bigr) \, \rho_m({\bm{r}})\ .
\label{Eq:MagMP}
\end{eqnarray}
  The partial integration of Eq. (\ref{Eq:MagMP}) leads to
\begin{eqnarray}
 \mathcal{M}_{\ell m} = \! \! \sqrt{\frac{4\pi}{2\ell +1}}\int d{\bm r}\, \nabla\Bigl(r^{\ell} Y_{\ell m}^{*}(\theta,\phi)\Bigr)\cdot {\bm
  m}({\bm r}) .
\label{Eq:MagMP2}
\end{eqnarray}
 This equation implies that the magnetic multipole moments are the
 quantities characterizing the angular distribution of the magnetization ${\bm m}({\bm r})$.
 Further details of this formulation can be found in Refs~\cite{blatt1952_book, kusunose2008}.


   The multipole moment is thus a quantity which classifies the spacial and angular shape  of a charge or magnetization distribution.
    Multipole moments are denoted according to their rank $\ell$, such as $\ell=0$: monopole, $\ell=1$: dipole, $\ell=2$: quadrupole, $\ell=3$: octupole, $\ell=4$: hexadecapole, $\ell=5$:  triakontadipole or dotriacontapole, and $\ell=6$:
 tetrahexacontapole. This nomenclature stems from the Greek notation for the $2^{\ell}$-numbers.
 The prefactors in Eqs.\ (\ref{Eq:ElecMP}) and (\ref{Eq:MagMP}) are chosen such that $\mathcal{Q}_{00}$ corresponds to
 the total enclosed charge and $\mathcal{M}_{1m}$ corresponds to the magnetic dipole moment, respectively, though the
 choice has arbitrariness.
   Electric multipole moments have the unit of
   $e$ $\times $ (length)$^{\ell}$ and magnetic multipoles have  $\mu_B \times$ (length)$^{\ell -1}$,
  where $e$ is the elementary charge and $\mu_B$ is the Bohr magneton.
   It is straightforward to see that the rank zero electric multipole moment corresponds to the enclosed total electric
   charge. The  ordinary magnetic moment is obviously given by  the
 rank-1 magnetic dipole moment. 
  The rank zero magnetic multipole moments, corresponding to magnetic
 monopole charge, do not appear in the multipole expansion, reflecting
 the relation $\nabla\cdot{\bm B}=0$,  but it 
   can be a useful concept for composite objects (see e.g.\ \cite{castelnovo2008,spaldin2008,khomskii2012,spaldin2013}).


A useful connection between multipole expressions using spherical harmonics and those based on Cartesian coordinates is obtained from
\begin{eqnarray}
 r^{\ell} Y_{\ell m}(\hat{{\bm r}})\!\! \!&=&\!\!\! \sqrt{\frac{2\ell + 1}{4\pi}(\ell +m)!(\ell -m)!}  \times \nonumber \\
   &&  \sum_{pqs}\biggl(-\frac{x+iy}{2}\biggr)^p\biggl(\frac{x-iy}{2}\biggr)^q z^s\ ,
\end{eqnarray}
 where $p$, $q$, and $s$ are zero or positive integers that satisfy $p+q+s=\ell$ and $p-q=m$.~\cite{varshalovich1988_book}


\subsection{Symmetry properties of multipole moments}

In condensed matter physics, multipole moments are often introduced as
mathematical tool to characterize the point group symmetry breaking. 
 The $\rho_e ({\bm r})$ and $\rho_m ({\bm r})$ are even and odd,
respectively, for the time reversal operation, leading to
$\mathcal{Q}_{\ell m}\rightarrow \mathcal{Q}_{\ell m}$ and
$\mathcal{M}_{\ell m}\rightarrow -\mathcal{M}_{\ell m}$ for the time
reversal operation.
 The space inversion operation transforms the electric charge density, magnetic
 charge density and magnetic moment density as: $\rho_e ({\bm r}) \rightarrow
 \rho_e (-{\bm r})$, $\rho_m ({\bm r}) \rightarrow
 -\rho_m (-{\bm r})$, ${\bm m} ({\bm r}) \rightarrow {\bm m} (-{\bm
 r})$, respectively. 
As a result, the relation $Y_{\ell m}(-{\hat{\bm
 r}})=(-1)^{\ell}Y_{\ell m}(\hat{{\bm r}})$ leads to the parity
 of the rank $\ell$ multipole moments for the spacial inversion as
 $(-1)^{\ell}$ for the electric multipoles and $(-1)^{\ell-1}$ for the
 magnetic multipoles.
 From these relations, the electric (magnetic) multipole moments are finite only
 for even (odd) rank when the system has the spacial inversion symmetry.

  The multipole moments are classified according to the irreducible
 representations (IREP) of the point group, taking a linear combination
 of Eqs.\ (\ref{Eq:ElecMP}) and (\ref{Eq:MagMP}) that is reflecting the
 transformation property under the corresponding point group operations.~\cite{shiina1997,kiss2003,kiss2005,takimoto2006b,kusunose2008}
 %
%
 To use this, it is convenient to start with considering the multipole
 moments classified according to the highest point group symmetry, i.e.\
 cubic group $O_{h}$ or hexagonal group $D_{6h}$. 
   The multipole expression can then be applied to multipole degree of
 freedom on an atomic site whose symmetry is lower than $O_{h}$ or
 $D_{6h}$, and the multipole moments are re-classified according to the
 IREPs of the point group.
In the following we adopt the Bethe notation, such as $\Gamma_7$
 and $\Gamma_8$, for the IREPs of orbitals
 including the effect of spin-orbit coupling and  Mulliken notation,
 such as $A_1$ and $E$, for
 the IREPs to which multipole moments belong, with the
 notation of $g$ and $u$ for even and odd parity for space inversion symmetry, respectively,
 and $+$ and $-$ for time reversal symmetry.

\section{Multipolar order}
\label{Sec:MPorder}
\subsection{First-principles approach to multipolar ordered phase}
\label{Sec:Method}

Multipolar order has been observed particularly in materials containing
elements with open $4f$ or $5f$ electron
shells~\cite{bucher1974,erkelens1987,kitagawa1996,lingg1999,yamauchi1999,tanaka1999,hirota2000,matsuoka2001,nakao2001,tayama2001,iwasa2002,paixao2002,hao2003,kohgi2003,aoki2005,tokunaga2005,onimaru2005,kiss2006,masaki2006,goto2009,carretta2010,caciuffo2011,onimaru2011,onimaru2016a,onimaru2016}. These
$f$ electron materials are  characterized by relatively localized and
correlated $f$ states, having a significant orbital disproportionation
and/or anisotropic hybridization, and a strong spin-orbit interaction. More than a decade
of research devoted to determining the optimal \textit{ab initio}
computational approach showed that good descriptions could be achieved
with the DFT+$U$ (Ref.\ \cite{anisimov1997}) and DFT+DMFT
methods~\cite{kotliar2006}.
 The $+U$ approach is particularly suitable for capturing the strong local
$f$-electron correlations whereas the DMFT approach is able to include
effects of low-energy dynamic fluctuations in the electronic structure,
beyond the Kohn-Sham DFT formulation.

 Both the LDA+$U$ and GGA+$U$ methods with S-O interaction included have been successful
  in capturing the ground state properties of $f$-electron compounds.~\cite{shick2001_1, shick2001_2, harima2001, harima2002, ghosh2005, shorikov2005, shick2005, shick2005b, shick2006, mtsuzuki2010, mtsuzuki2010_2,soderlind2010} In spite of some limitations this method is especially suited to
  describe the local character of $f$ electrons on the same footing with
  the electronic band description.
The LDA+$U$ method~\cite{anisimov1991, czyzyk1994, anisimov1997, shick1999} provides the one-electron Hamiltonian as
\begin{eqnarray}
h_{{\rm LDA}+U}=h_{\rm LDA}+\sum_{\tau}\sum_{\gamma\gamma'}|\tau\ell\gamma\rangle v_{\gamma\gamma'}^{\tau\ell} \langle \tau\ell\gamma'|\ ,
\label{Eq:oneHamilton}
\end{eqnarray}
where $h_{\rm LDA}$ is the conventional Kohn-Sham single-electron
 Hamiltonian that contains the kinetic energy, Coulomb-Hartree
 interactions, exchange-correlation energies, and relativistic
 correction terms, $|\tau\ell\gamma \rangle$ denote the local basis set,
 $\gamma$ ($\gamma'$) is an index related to an orbital $m$
 ($m^{\prime}$) and a spin $s$ ($s'$) quantum number, or, alternatively, double-valued
 irreducible representations of the site symmetry of the correlated $f$
 electron  site obtained through a unitary transformation.
 The on-site Coulomb potential is given by
\begin{eqnarray}
\!\! \!\! v_{\{ms\}\{m's'\}}^{\tau\ell} &=&\sum_{m''m'''} \Big[ \delta_{ss'} \sum_{s''}n^{\tau\ell}_{\{m'''s''\}\{m''s''\}} \times \nonumber \\
& & \langle mm''|W|m'm'''\rangle \nonumber \\
&-&n_{\{m'''s\}\{m''s'\}}^{\tau\ell}\langle mm''|W|m'''m'\rangle \Big] \nonumber \\
&-&\delta_{mm'}\delta_{ss'} \big[ U(n^{\tau\ell}-\frac{\eta^{\tau\ell}}{2})-J(n^{\tau\ell}_{s}-\frac{\eta_{s}^{\tau\ell}}{2})\big] ,
\label{eq:Upotential}
\end{eqnarray}
where $\tau$ and $\ell$ are the atom's index and angular momentum of
the orbitals, respectively, for which the $+U$ potentials are
introduced.
$n^{\tau\ell}_{\{ms\}\{m's'\}}$ is the local spin-orbital electron occupation
matrix,
$n_{s}^{\tau\ell}=\sum_{m=-\ell}^{\ell}n^{\tau\ell}_{\{ms\}\{ms\}}$,
$n^{\tau\ell}=\sum_{s}n^{\tau\ell}_{s}$. $\eta^{\tau\ell}=\frac{1}{2}\sum_{s}\eta_{s}^{\tau\ell}$
depends on the type of double counting term, specifically, $\eta_{s}^{\tau\ell}=1$ for the
fully localized limit and
$\eta_{s}^{\tau\ell}=\frac{1}{2\ell+1}\sum_{m=-\ell}^{\ell}n^{\tau\ell}_{\{ms\}\{ms\}}$
for the around mean field formulation~\cite{anisimov1993,czyzyk1994}.
 The electron occupation matrix $n^{\tau\ell}_{\gamma\gamma'}$ pertaining to a certain ion can be calculated for the local space spanned by the local
bases inside the atomic or muffin-tin (MT) spheres around the selected
atom as:
\begin{eqnarray}
n_{\gamma\gamma'}^{\tau\ell} = \int_{\rm MT} dr_{\tau} r_{\tau}^2 \, 
 \rho^{\tau\ell}_{\gamma\gamma'}(r_{\tau})\ , 
\label{Eq:occupationmat}
\end{eqnarray}
with
\begin{eqnarray}
\rho^{\tau\ell}_{\gamma\gamma'}(r_{\tau})= \frac{1}{N}
 \sum_{\boldsymbol{k}b}\langle \tau\ell\gamma | \boldsymbol{k}b \rangle
 \ \langle \boldsymbol{k}b | \tau\ell\gamma'\rangle \ ,
\label{Eq:densitymat}
\end{eqnarray}
where $| \boldsymbol{k}b \rangle$ are the Bloch band states (eigenstates of Eq.\ (\ref{Eq:oneHamilton})) that are projected on to the local basis,
$N$ is the number of $\boldsymbol{k}$ points in reciprocal space and $r_{\tau}$ is the radial component of the position vector
$\boldsymbol{r}_{\tau}$ measured from the center of the atom $\tau$.
The occupation matrix as well as the charge
density is determined selfconsistently in the framework of the LDA+$U$ method.
The matrix elements of the Coulomb interaction of $f$ electrons is expressed as
\begin{eqnarray}
\langle m_1m_2\mid W \mid m_3m_4\rangle =\delta_{m_1+m_2,m_3+m_4}\nonumber\\
~~ \times\sum_{k=0}^{6}c^k (\ell m_1;\ell m_3)c^k(\ell m_4;\ell m_2)F^k\ ,
\end{eqnarray}
where the $F^k$ are the Slater-Condon parameters,~\cite{slater1929,condon1931} and $c^k$ is the Gaunt coefficient.~\cite{gaunt1929,racah1942}
In practical calculations $F_0$ is taken as $F_0=U$, the Hubbard $U$
parameter. The Hund's coupling parameter $J$ is related with the higher
order Slater integrals as $J=(286F_2+195F_4+250F_6)/6435$ for $f$
electrons~\cite{anisimov1997}.
 The ratio between the higher order Slater integrals is obtained from
hydrogenic radial wave functions, such as $F_4/F_2=0.138$ and
$F_6/F_2=0.0151$~\cite{judd1955}, and it is convenient to choose only $U$ and $J$
as parameters in the LDA+$U$ calculations.
 
  The multipolar ordered states are calculated by introducing the specific
  symmetry breaking by considering the local symmetry of the multipole
  moment, which is characterized by the IREP, and the
  configuration of the multipole moments in the crystal
  structure. General symmetry properties of multipolar ordered
  states are discussed in the Appendix.
 In the LDA+$U$ method, one can introduce the symmetry breaking through the initial electron
  occupation matrix, since appearing multipole order parameters involve
  the multiple spin and orbital degrees of freedom as well as the local
  occupation matrix, that make it suitable to calculate
  complex multipolar ordered phases. 
 For instance, nonmagnetic calculations are performed with the relation
 $n_{\{-m-s\}\{-m'-s'\}}^{\tau\ell}$=$(-1)^{m+m'+s-s'}$$n_{\{ms\}\{m's'\}}^{\tau\ell*}$
 for the electron occupation matrix, Eq.\  (\ref{Eq:occupationmat}),
 to  preserve the time reversal symmetry~\cite{mtsuzuki2010_2}, and this
 relation should be removed to calculate the magnetic multipolar ordered
 states. Other relations between the matrix elements are identified by
 investigating the transformation property of the density matrix for the
 point group operations, depending on the local atomic site symmetry on which the density matrix
 is defined. Furthermore, the relation between the local principal axes of the multipole
 moments on different atoms are determined when the magnetic space group
 of the ordered state is identified (See the Appendix).
Since a large Coulomb $U$
 tends to increase the anisotropic character of the $f$ states, the
 full-potential treatment~\cite{weinert1980} is an important ingredient, too, to adequately
 reproduce the behavior of anisotropic $f$ states in correlated materials.
In recent explicit calculations  the full-potential linearized augmented
 plane wave (FLAPW) band-structure method with the +$U$ implementation has therefore been used~\cite{mtsuzuki2010_2,mtsuzuki2010,mtsuzuki2013,mtsuzuki2009}.

It is known that the large $U$ introduced in the LDA+$U$ method can induce some meta-stable states especially in calculations of ordered states and may lead to convergence to an electronic state that is inconsistent with the realistic ground state.~\cite{shick2001_2, larson2007, jomard2008, dorado2009, dorado2011}
 To avoid this problem, experimental information concerning the CEF
 ground states and the order parameters can be used to control the occupations for the initial density matrix and guide convergence to the  proper ground state.
 
The local multipole moments can be obtained from the local
 spin-orbital occupation matrix. This quantity is
 computed selfconsistently, starting from an initial density matrix and allowing for suitable
symmetry breaking.
The expectation values of the local operators $O^{\tau\ell}$
defined for the local basis on a specific atom $\tau$ and orbital $\ell$ are calculated with the local basis set \{$|\tau\ell\gamma \rangle$\} inside the MT spheres, following the expressions
\begin{eqnarray}
 O^{\tau\ell}( \boldsymbol{r}_{\tau}) &\equiv& \frac{1}{N^2}\sum_{\boldsymbol{k} b}\sum_{\boldsymbol{k}' b'}\sum_{\gamma\gamma'} \langle \boldsymbol{r}_{\tau} | \boldsymbol{k} b \rangle \ \langle \boldsymbol{k} b | \tau\ell\gamma\rangle O^{\tau\ell}_{\gamma\gamma'} \nonumber \\
&&\times \langle \tau\ell \gamma' | \boldsymbol{k}' b' \rangle \ \langle \boldsymbol{k}' b' | \boldsymbol{r}_{\tau}\rangle ,  
\label{Eq:MPDist}
\end{eqnarray}
and the integration inside the muffin-tin sphere is 
\begin{eqnarray}
\!\!\! \!\! \langle O^{\tau\ell}\rangle &=& \!\!\!
\int_{\rm MT} d \boldsymbol{r}_{\tau} O^{\tau\ell}( \boldsymbol{r}_{\tau})  \nonumber \\
&=& \!\!\!\sum_{\gamma} \sum_{\gamma_1\gamma_2} \int \! dr_{\tau} r_{\tau}^{2}\, \rho^{\tau\ell}_{\gamma\gamma_1}(r_{\tau})O^{\tau\ell}_{\gamma_1\gamma_2}\rho^{\tau\ell}_{\gamma_2\gamma}(r_{\tau}) . 
\label{Eq:MPExpect}
\end{eqnarray}
 A matrix element of the multipole operators is systematically
 calculated with Steven's operators technique, and 
the explicit expressions for the local multipole operators $O^{\tau\ell}$
for sites with specific local crystal symmetries have been
listed~\cite{shiina1997,kusunose2008}.
 When the multipole moments align
 antiferromagnetically, the antiferromagnetic alignment of the local principal axis at
 each atomic site is assured by posing the proper magnetic space group symmetry
 for the antiferroic multipolar order.

\subsection{ Introduction to the $An$O$_2$ compounds}
\label{Sec:IntroAnO2}
 Recently, $f$ electron materials containing actinide elements have
 drawn considerable interest, stimulated by observations of intriguingly
 ordered ground states emerging at low temperatures. These conditions are
 in particular met in the rare-earth and actinide compounds, which 
have provided a treasure trove of a rich variety of multi-orbital
 physics over many years.
 A striking example is the low-temperature ordered ground state of
 NpO$_2$, which, after many years of investigations could experimentally
 be established to be due to a high-rank magnetic multipolar order, in
 the absence of any dipolar moment formation.\cite{kuramoto2008,
 kuramoto2009,santini2009}

 The quest for the hidden order parameter in NpO$_2$ started sixty years ago when an unusual phase transition
to an unknown  ordered phase was discovered below $T_0 \sim 25$ K in specific heat measurements.~\cite{osborne1953} 
Subsequent magnetic susceptibility measurements revealed also a clear phase-transition anomaly\cite{ross1967,erdos1980},
but no ordered dipole magnetic moment could be detected in neutron scattering experiments \cite{cox1967,caciuffo1987}
and M{\"o}ssbauer spectroscopy.~\cite{dunlap1968, friedt1985} However, muon spin rotation ($\mu$SR) measurements revealed
the breaking of time-reversal symmetry in the mysteriously  ordered phase. \cite{kopmann1998}
Tremendous experimental and theoretical efforts have been invested to reach an explanation of the peculiar low-temperature phase (see e.g.\ Ref.\cite{santini2009}). A higher order magnetic octupole moment was first suggested. \cite{santini2000,santini2002}
Important insight was obtained from resonant x-ray scattering~\cite{mannix1999} that identified the electric symmetry as triple-${\bm q}$ antiferro quadrupolar (AFQ) long range order of (111) oriented multipole moments below $T_0$  and suggested the (time-reversal symmetry broken) antiferro order of $T_{2g}$-magnetic multipole moments as primary order parameter~\cite{paixao2002, caciuffo2003}. X-ray Bragg scattering experiments were 
 initially interpreted as evidence for the absence of rank-5 multipole
 moments\cite{lovesey2003}, but this possibility was
re-examined later~\cite{lovesey2012}.
  Nuclear magnetic resonance (NMR) experiments further supported the 3${\bm q}$ ordering of magnetic multipole moments, detecting the splitting of the $^{17}$O spectra due to 
the symmetry lowering around the oxygen sites accompanied by the 3${\bm
 q}$ ordering below $T_0$~\cite{tokunaga2005, tokunaga2006, walstedt2007}.
Then, it has been realized that the order parameter has $T_{{\rm 2}g}$ symmetry with type-I AF ordering of magnetic multipoles.~\cite{sakai2003,sakai2005}
 Inelastic neutron scattering measurements and theoretical analysis could finally identify triple-${\bm q}$ antiferro ordered multipoles  of  rank-5 (triakontadipoles) of
 the $T_{{\rm 2}g}$ symmetry as the leading order parameter.~\cite{santini2006,magnani2008, mtsuzuki2010}

\begin{figure}[t]
	\includegraphics[width=1.0\linewidth]{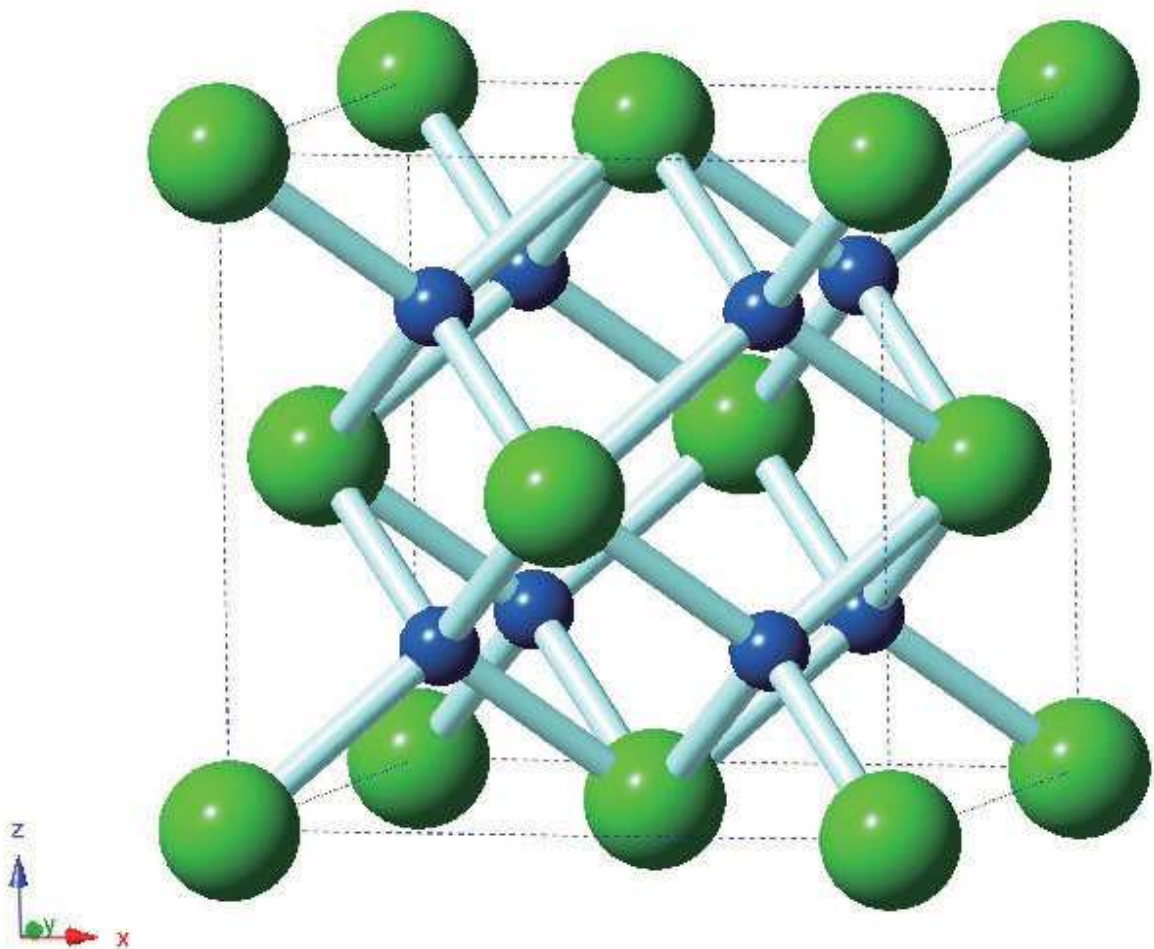}
	\caption{ (Color online) The cubic fluorite crystal structure of the $An$O$_2$
 compounds. Green spheres depict the actinide $An$ ion, blue spheres the
 oxygen atoms.}
\label{Fig:AnO2Cryst}
\end{figure}

\begin{figure}[t]
	\includegraphics[width=1.0\linewidth]{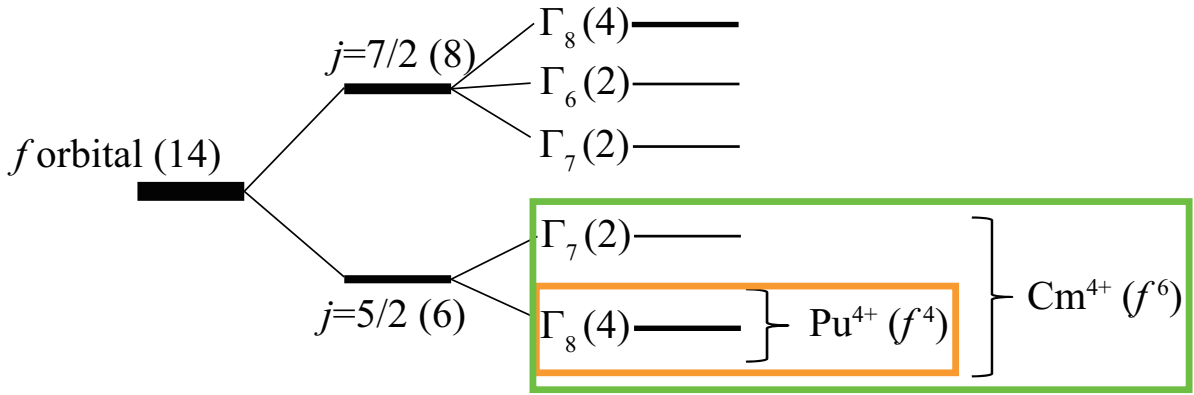}
	\caption{ (Color online) Schematics of the splitting of the 14
 one-electron $f$ orbitals. Spin-orbit interactions splits the orbitals
 in $j=5/2$ and $j=7/2$ orbitals, which are further split by the cubic
 crystal field. The number in the brackets denotes the degeneracy of the
 orbitals (from Ref.\ \cite{mtsuzuki2013}).
\textcopyright 2013 American Physical Society}
	 	\label{Fig:f_orbital_split}
\end{figure}

 While the physical features of NpO$_2$ are outstanding, the other actinide dioxides are of scientific interest as well~\cite{santini1999}.
All known $An$O$_2$ compounds with $5f$ electrons ($An$ = U, Np, Pu, Am,
and Cm) crystallize in the FCC structure, shown in Fig.\ \ref{Fig:AnO2Cryst}, characterized 
by the space group $Fm\bar{3}m$ ($O_h^5$, No.\ 225), and all compounds are insulators with sizeable band gaps of about 2 eV~\cite{mcneilly1964,schoenes1980,mccleskey2013}.
 The 14 one-electron $f$ orbitals are split by the
spin-orbit interaction in the $j=5/2$ and $j=7/2$ orbitals, which are
further split by the cubic CEF, see Fig.\ \ref{Fig:f_orbital_split}. 
On the basis of the point charge model, the energy level of the
$\Gamma_{8}$ one-electron orbital is expected to be lower than that of
the $\Gamma_{7}$ orbital in the $j=5/2$ orbitals since the oxygen anions
are located along the [111] direction, to which the $\Gamma_7$ orbital
extends~\cite{kubo2005c, kubo2005b}.
The filling of the $\Gamma_{8}$ orbital by two, three, and four $f$
electrons leads to the $\Gamma_5$ triplet, $\Gamma_8^{(2)}$ quartet, and
$\Gamma_1$ singlet CEF ground state in the \textit{L-S} coupling
scheme, which are consistent with the experimental observations for
UO$_2$~\cite{kern1985,amoretti1989},
for NpO$_2$~\cite{fournier1991,amoretti1992}, and for
PuO$_2$~\cite{kern1990,kern1999}, respectively.

In spite of the alike insulating and structural properties, the $An$O$_2$ compounds show a wide variety of magnetic ground states. 
UO$_2$ exhibits an antiferromagnetically ordered ground state below $T = 30$ K~\cite{osborne1953,willis1965,frazer1965,faber1975,faber1976} 
that later on was shown to be a 3$\bm q$ magnetic dipole ordered state~\cite{ikushima2001,wilkins2004,wilkins2006}.
The dipolar 3${\bm q}$ magnetic order is accompanied by a 3${\bm q}$ ordering of quadrupoles, which in turn are related  to a distortion of the 5$f$ charge density in the direction of dipole moments \cite{carretta2010,wilkins2006}. 

PuO$_2$ does not show any magnetic phase transition, rather, the magnetic susceptibility is temperature-independent 
up to 1000~K \cite{raphael1968}. The nonmagnetic ground state of PuO$_2$
has been confirmed by a recent Pu-NMR study \cite{yasuoka2012} and it
can be understood from a CEF analysis for the Pu$^{4+}$ ion that gives a
nonmagnetic singlet state as the CEF ground state \cite{kern1990,kern1999,santini1999}.

AmO$_2$ undergoes a conspicuous phase transition at around
 8.5~K.\cite{karraker1975} While a peak structure in the magnetic
 susceptibility was found,\cite{karraker1975} neutron-diffraction
 measurements could not detect any antiferromagnetic dipolar order in agreement with the M{\"o}ssbauer measurement.\cite{kalvius1969, boeuf1979} Hence, the features of AmO$_2$ are similar to those
 of NpO$_2$, and therefore an antiferro multipolar ordered ground state is expected in AmO$_2$.
The CEF ground state of the Am$^{4+}$ ion in AmO$_2$ was thought to be a $\Gamma_7$ doublet.~\cite{karraker1975, abraham1971, kolbe1974}
However, the $\Gamma_7$ state has no degree of freedom for the higher
 rank multipoles and therefore seems to contradict the experiments. A
 recent CEF analysis based on the $j$-$j$ coupling model discussed an
 instability of the $\Gamma_7$ ground state and possibility of
 stabilization of the $\Gamma_8$ ground state, which could induce higher
 order multipoles without inducing a dipole moment.\cite{hotta2009} 
 Notably, there exist many experimental challenges to distinguish the essential bulk contribution, because of the strong self-radiation damage caused by alpha decay in this material.\cite{benedict1980, edelstein2006, tokunaga2010, tokunaga2011}

For the rare human-made compound CmO$_2$ only a few experiments have thus far been reported.\cite{morss1989,kvashnina2007}  The Cm$^{4+}$ ion ($5f^6$ occupancy) is expected to have a nonmagnetic ground state
(Fig.\ \ref{Fig:f_orbital_split}), however a paramagnetic moment has been detected \cite{morss1989}; This observation has been explained within the $j-j$ coupling model assuming the energy proximity of a magnetic excited state~\cite{niikura2011}.

The anomalous low-temperature properties of the $An$O$_2$ compounds have been theoretically addressed 
with various approaches.  CEF analyses have been carried out for UO$_2$ (Refs.\ \cite{rahman1966,kern1985,amoretti1989,zhou2011}), for NpO$_2$ (Refs.\ \cite{fournier1991,amoretti1992}), and the whole series \cite{magnani2005}. Theory of the 3$\bm q$ magnetic order of UO$_2$ has also been developed\cite{allen1968_1,allen1968_2,burlet1986,giannozzi1987,mironov2003}, as well as $j-j$ coupling theory \cite{kubo2005a,kubo2005b} and multipole symmetry analysis \cite{matteo2007} for NpO$_2$.
First-principles theory has also been employed to unravel physical and chemical properties of the $An$O$_2$ compounds
\cite{brooks1983,kudin2002,colarieti2002,laskowski2004,prodan2006,prodan2007,yu2011,yin2011,wen2012}
The vibrational lattice dynamics measured by inelastic neutron or inelastic x-ray scattering could recently be explained by 
GGA+$U$ calculations for UO$_2$ and NpO$_2$ \cite{pang2014,maldonado2016}.
However, first-principles calculations of the ordered multipolar phases is an entirely different matter, as outlined in the following.

\subsection{ First-principles calculation of $An$O$_2$ ground states}
\begin{figure}[t]
	\includegraphics[width=1.0\linewidth]{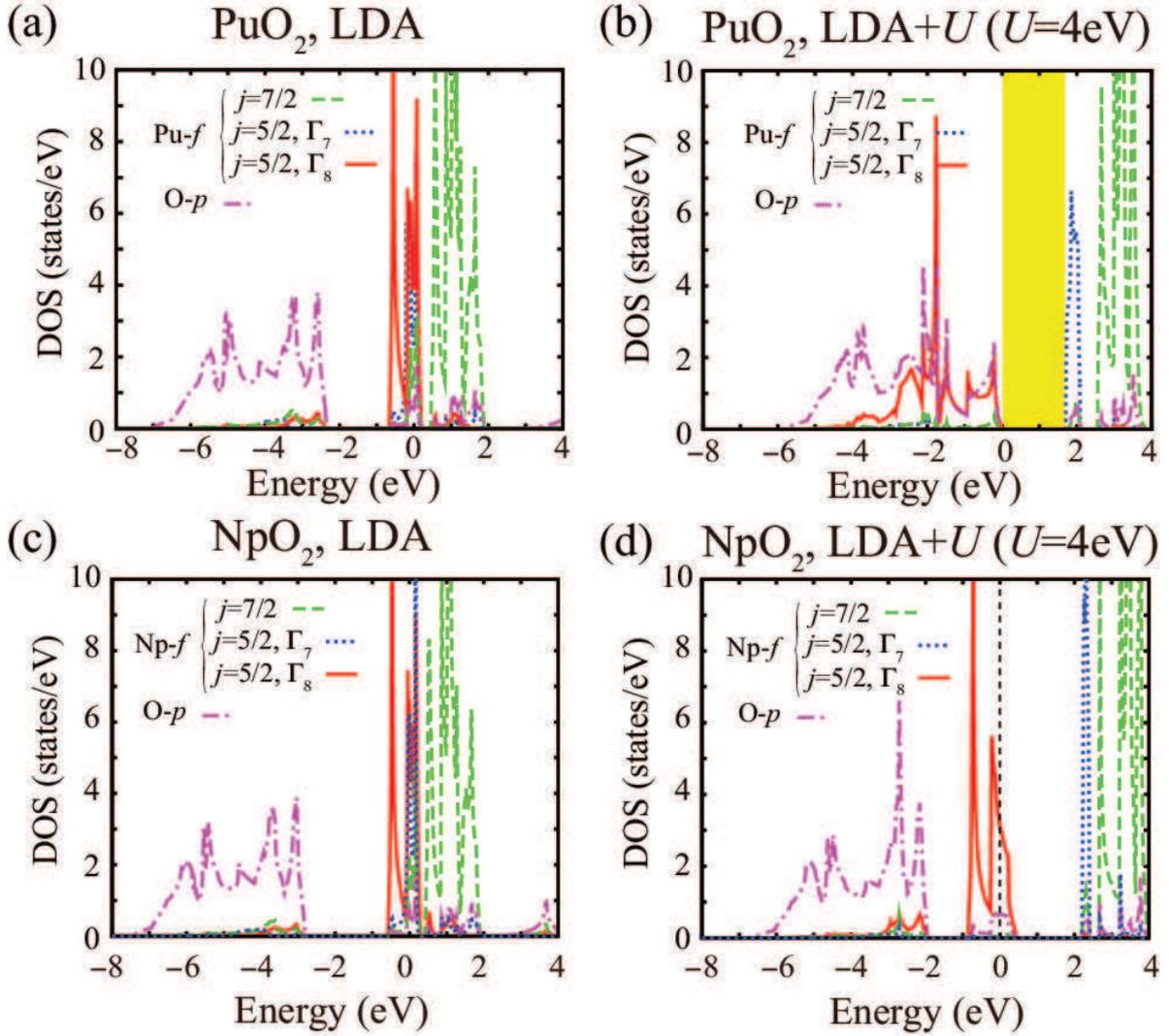}
	\caption{ (Color online) Density of states (DOS) calculated for the nonmagnetic
 solutions of PuO$_2$ and of NpO$_2$, left panels using the LDA
 and right panels using the LDA$+U$ method.  The $j=5/2$
 orbitals are projected on the $\Gamma_7$ and
 $\Gamma_8$ one-electron orbitals shown in Fig.\ \ref{Fig:f_orbital_split}. From Ref.\ \cite{mtsuzuki2013}} 
	\label{Fig:DOSAnO2_Nonmag}
\end{figure}
\begin{figure}[t]
\begin{center}
	\includegraphics[width=0.8\linewidth]{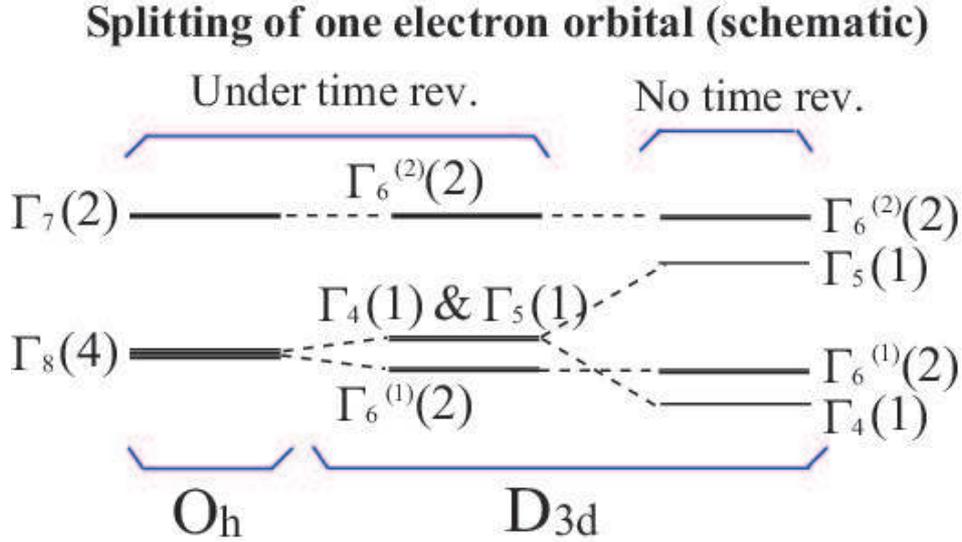}
		\end{center}
	\caption{ (Color online) One electron energy level scheme of the Np $j=5/2$ $f$
 orbitals at the local Np site in NpO$_2$. The $\Gamma_8$ quartet in the cubic $O_h$ symmetry
 with time-reversal symmetry splits in the reduced $D_{3d}$ symmetry ($Pn\bar{3}m$ space group) 
 in one doublet $\Gamma_6^{(1)}$, and two  singlets, $\Gamma_4^{ }$ and
 $\Gamma_5^{ }$ that are degenerate under time-reversal symmetry. This
 degeneracy is lifted by time-reversal symmetry breaking. From Ref.\
 \cite{mtsuzuki2013}
\textcopyright 2013 American Physical Society} 
	\label{Fig:LevelScheme_NpO2}
\end{figure}
\begin{figure}[t]
	\includegraphics[width=0.8\linewidth]{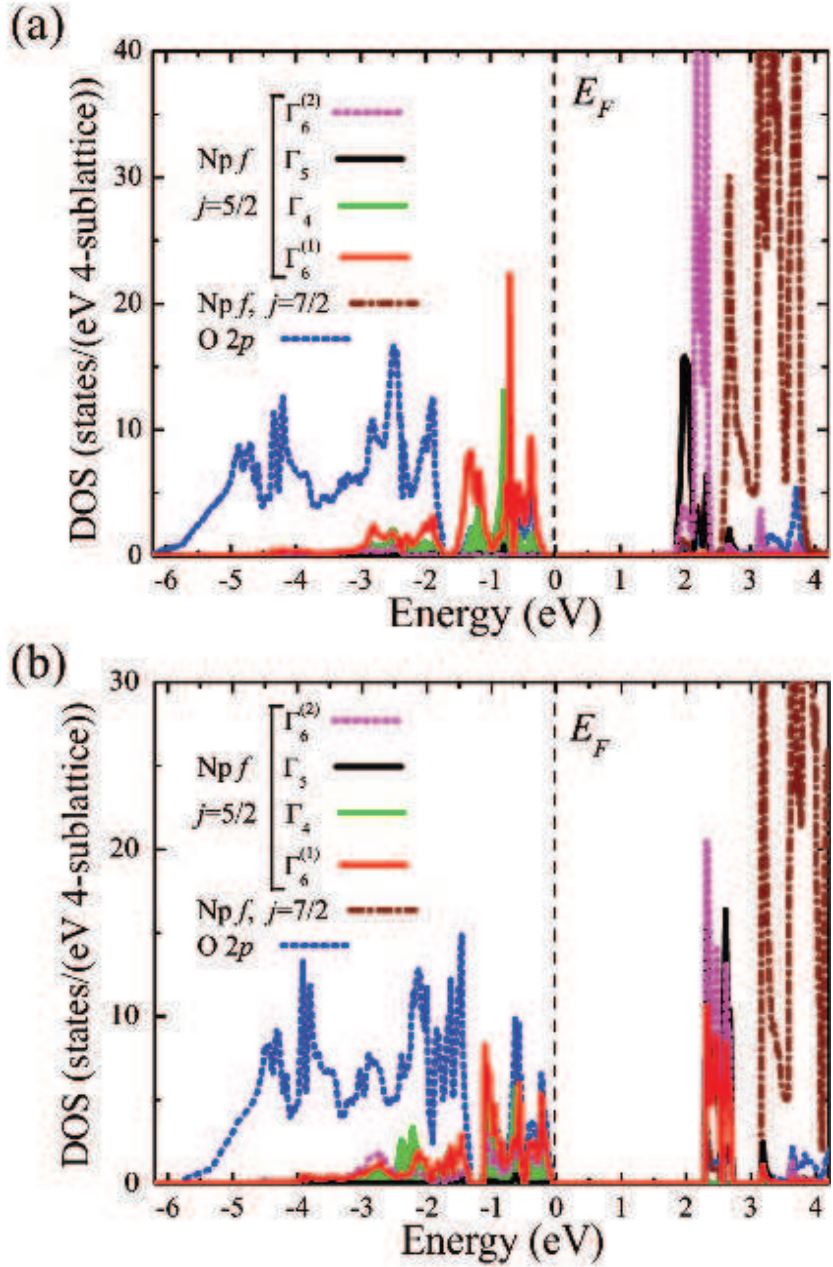}
	\caption{ (Color online) Density of states calculated for the magnetic 3${\bm q}$ $T_{2g}$ multipolar
 ordered state of NpO$_2$ for $U=4$\,eV, (a) with $J=0$ and (b) with $J=0.5$\,eV. The Np $j=5/2$ states are projected
 on the one-electron orbitals given schematically in Fig.\
 \ref{Fig:LevelScheme_NpO2}.~\cite{mtsuzuki2010,mtsuzuki2013}
\textcopyright 2010 The American Physical Society} 
	\label{Fig:DOSNpO2_Order}
\end{figure}
\begin{figure}[t]
\begin{center}
	\includegraphics[width=0.8\linewidth]{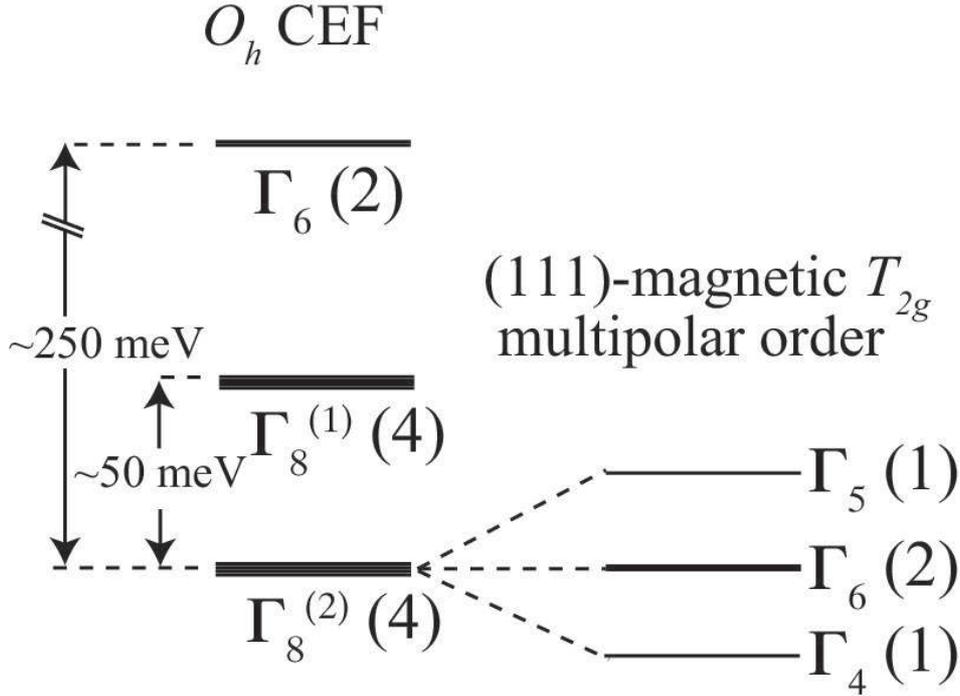}
		\end{center}
	\caption{Paramagnetic $O_h$ CEF levels~\cite{amoretti1992,santini2006} and splitting of the
 ground-state $\Gamma_{8}$ quartet in the magnetic multipolar ordered
 state of NpO$_2$ within the \textit{L-S} coupling scheme.
}
	\label{Fig:CEF_Multipole_NpO2}
\end{figure}
\begin{figure}[t]
\begin{center}
	\includegraphics[width=0.8\linewidth]{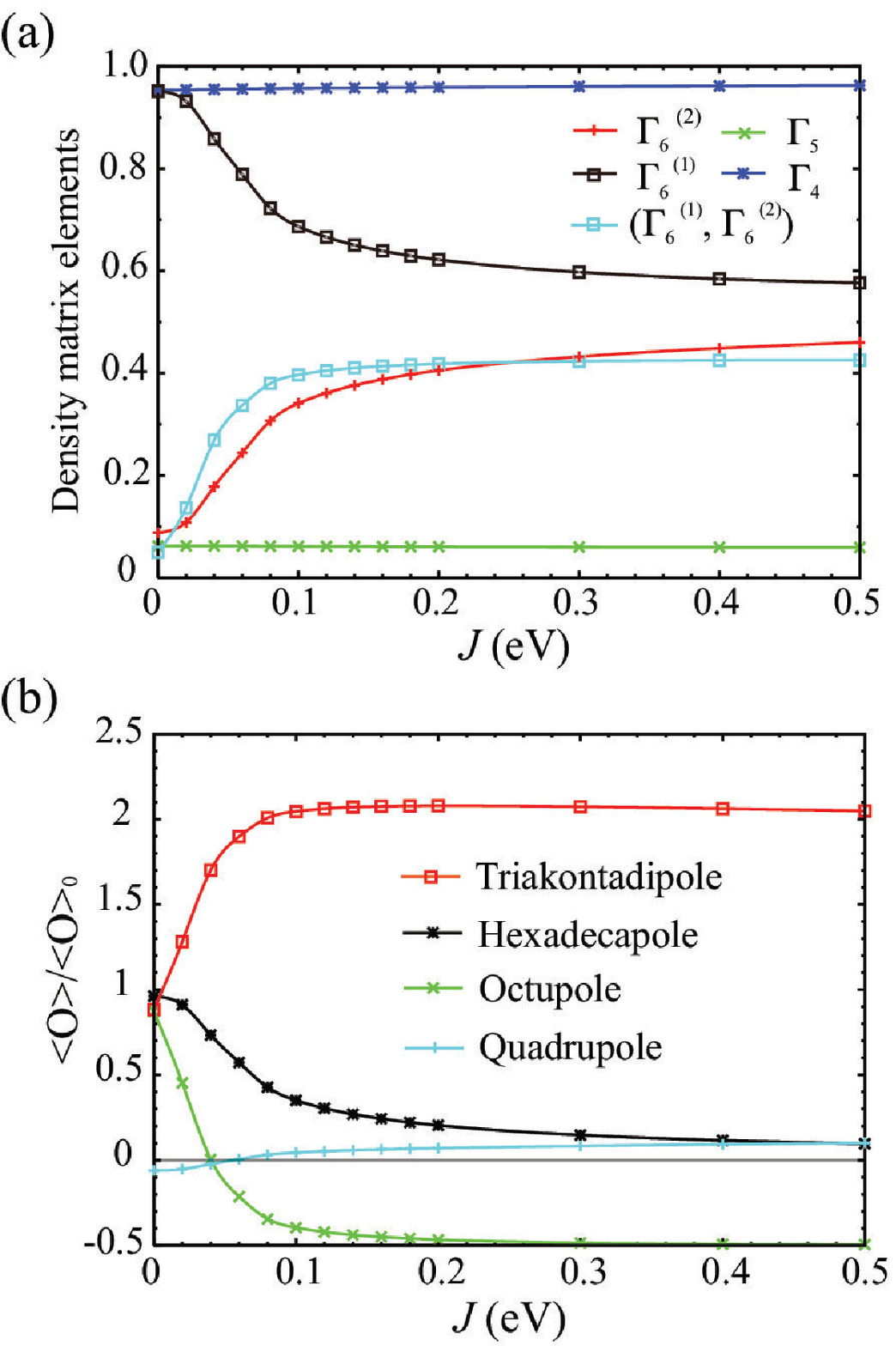}
	\caption{ (Color online) Calculated dependence on the Hund coupling parameter
 $J$ of (a) the electron occupation matrix elements, and (b)
	 the multipole moments of NpO$_2$~\cite{mtsuzuki2010,mtsuzuki2013}. 
\textcopyright 2010 The American Physical Society}
	\label{Fig:Multipole_NpO2}
\end{center}
\end{figure}

The electronic structure calculation is necessary to understand the origin of
 the energy gap formation in materials.
 However, the basic approximations of the first-principles calculation, such as the local density approximation (LDA) and
 generalized gradient approximation (GGA), do not reproduce the energy
 gap observed in the $An$O$_2$ compounds~\cite{petit1996,maehira2007,jomard2008,mtsuzuki2013,hasegawa2013}.
 First-principles calculations taking into account the strong Coulomb
 interaction have been applied to UO$_2$~\cite{brooks1983,kudin2002,laskowski2004,prodan2006,prodan2007,yu2011,zhou2011,wen2012},
 to NpO$_2$~\cite{mtsuzuki2010,prodan2007,wen2012}, and to
 PuO$_2$~\cite{prodan2006,prodan2007,wen2012,colarieti2002,petit2003,sun2008,jomard2008,modin2011,nakamura2011,wang2012}
 though some of the studies assume ordered states different from the
 experimentally observed ones.
Since the strong Coulomb interaction must be included in the calculations to
 reproduce the energy gap, the insulating $An$O$_2$ compounds are
 referred to as Mott insulators~\cite{laskowski2004, yin2008, yin2011}.

 The detailed electronic structure investigation for the ground states
 of $An$O$_2$ was performed using the LDA+$U$ method~\cite{mtsuzuki2010, mtsuzuki2013}.
In the LDA+$U$ calculations, the Coulomb $U$ parameter has been chosen as $U=4$ eV and the exchange $J$ in the range of $0  - 0.5$ eV. 
These values have previously been shown to provide an accurate description of measured properties of actinide dioxides.\cite{dudarev1997, dudarev1998, modin2011}
The double-counting term has been chosen as in the fully localized
 limit,\cite{liechtenstein1995} leaving out the spin dependency of the
 Hund's coupling part to adapt it for the nonmagnetic LDA part of Eq.\
 (\ref{eq:Upotential}).
  To illustrate the effect of strong Coulomb interaction on the energy gap formation
 we focus first on the electronic structures of PuO$_2$ and of NpO$_2$, and compare
  their ground state's electronic property, as obtained by 
 LDA and LDA+$U$ calculations.
   Figures \ref{Fig:DOSAnO2_Nonmag}(a) and (b) show the density of
  states {\ (DOS)} of PuO$_2$ obtained from LDA and LDA+$U$ calculations, respectively, assuming 
  the nonmagnetic constraint as discussed in Sec.\ \ref{Sec:Method}.  
 In the LDA calculation, the $\Gamma_7$ and $\Gamma_8$ $j=5/2$ orbitals of
 $f$ electrons are present in the narrow energy region around the Fermi energy resembling
 energetically degenerate sextet orbitals, due to the absence of orbital
  dependence of the Coulomb interaction. 
 Since the four $f$ electrons of the Pu$^{4+}$ ion occupy the sextet-like orbitals, the resulting
  electronic structure has the Fermi level located around the middle of the
  $j=5/2$ orbitals and hence results in a metallic state.
   Conversely, the LDA+$U$ calculations reproduce the energy gap by inducing the
  orbital splitting between the $\Gamma_7$ and $\Gamma_8$ orbitals with
  the full occupation of the $\Gamma_8$ quartet orbitals, corresponding
  to the $\Gamma_1$ singlet CEF ground state observed experimentally~\cite{kern1990,kern1999}.
  The nonmagnetic LDA+$U$ calculations induce a similar splitting for
 the $f$-orbitals of NpO$_2$ as shown in Figs.\ \ref{Fig:DOSAnO2_Nonmag}(c) 
 and (d). However, since the Np$^{4+}$ ion has three $f$
  electrons, the LDA+$U$ calculations lead to a metallic ground state
  within the nonmagnetic calculations due to the incomplete occupation
  of the $\Gamma_8$ quartet orbitals as shown in Fig.\
 \ref{Fig:DOSAnO2_Nonmag}(d).
 The large Coulomb repulsion in NpO$_2$ however renders the metallic
  states with the dense $f$-states around the Fermi level unstable and a certain
  magnetic order is expected to appear to obtain a total
  energy gain through the concomitant one-electron orbital splitting.
\begin{figure}[t]
\begin{center}
	\includegraphics[width=0.6\linewidth]{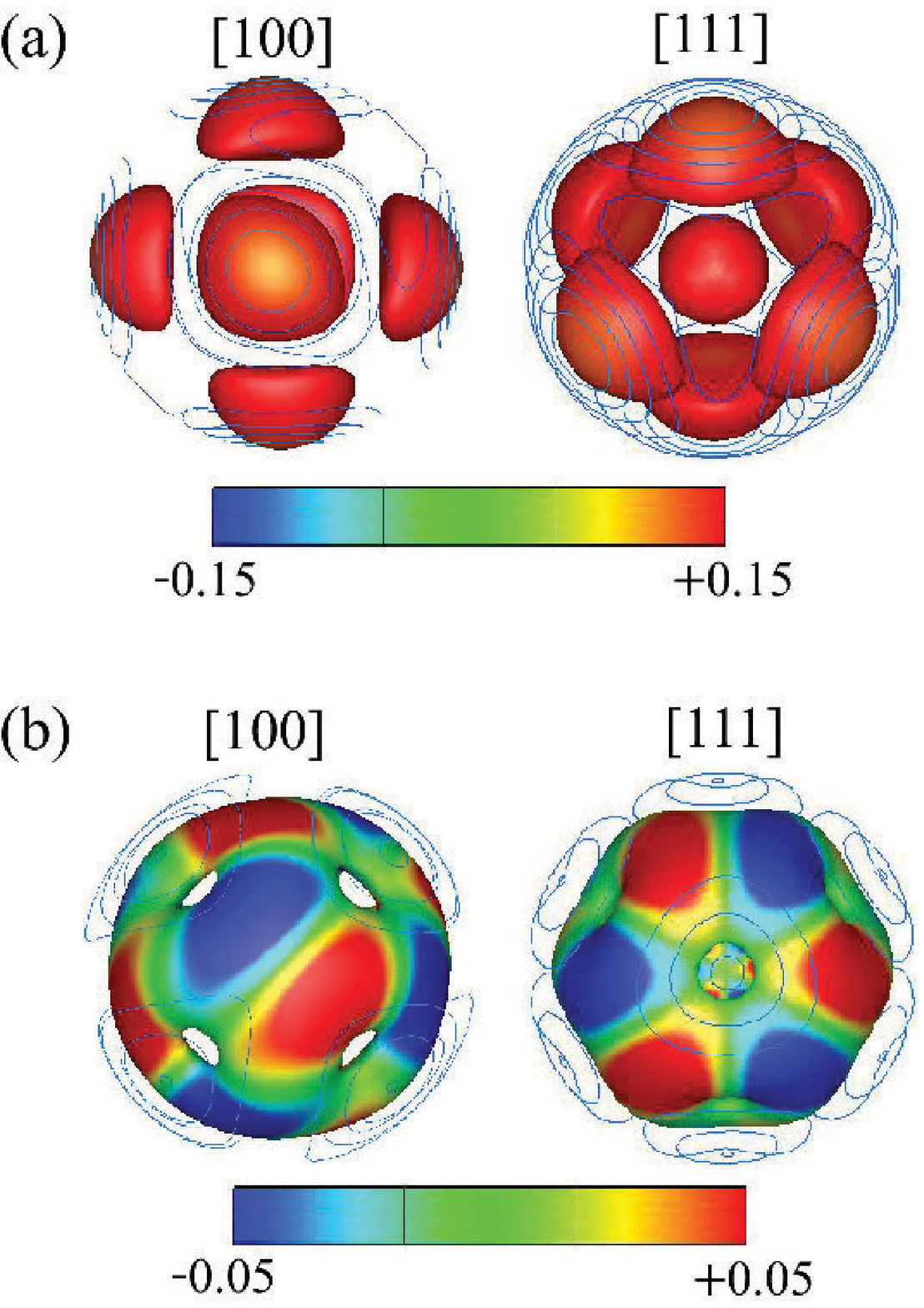}
	\caption{ (Color online) Spacial distributions of the  magnetic
 moment densities of (a) UO$_2$ and (b) NpO$_2$, for two
 different viewing directions, [100] and  [111], computed with $U=4$\,eV and $J =0.5$\,eV.  The magnetic moment
 distributions are depicted on the isosurfaces of the charge
 densities for the [111] component,
 with magnitudes as given by the color bars with $\mu_B$ unit.
  The thin lines show the contour map of the charge density on a
 spherical surface. From Ref.~\cite{mtsuzuki2013}
\textcopyright 2013 American Physical Society} 
	\label{Fig:ChgSpn_U_Np}
\end{center}
\end{figure}

  To obtain the 3${\bm q}$ antiferroic multipolar ordered state in selfconsistent
  electronic structures calculations the proper symmetry of the magnetic space group needs to be considered.
  Here, these require specifically a unit cell consisting of four elementary unit cells and 
 the magnetic space group $Pa\bar{3}$
 for UO$_2$ and $Pn\bar{3}m$ for
 NpO$_2$~\cite{nikolaev2003,mtsuzuki2010,mtsuzuki2013}.
 The influence of the reduced symmetry of the latter space group on the
 one-electron level scheme is shown in 
 Fig.\ \ref{Fig:LevelScheme_NpO2}. The reduced $D_{3d}$ symmetry leads
  to a splitting of the quartet $\Gamma_8$ one-electron orbital into two energy-degenerate singlet
  orbitals, $\Gamma_4$ and $\Gamma_5$, and a
  doublet $\Gamma_6^{(1)}$ orbital, in the absence of 
 time-reversal symmetry breaking. 
Allowing for time-reversal symmetry breaking leads to a further
  splitting of the two degenerate singlet orbitals. 
  Hence, the magnetic $T_{2g}^{-}$ multipolar order of NpO$_2$ could split the $j=5/2$
  $f$ orbitals as shown in Fig.\ \ref{Fig:LevelScheme_NpO2}. 
   Adopting the $D_{3d}$ point group symmetry and time-reversal symmetry breaking,
  the LDA+$U$ calculation indeed predicts the insulating ground state
  with the characteristic energy gap of around
  2\,eV as shown in Fig.\ \ref{Fig:DOSNpO2_Order}~\cite{mtsuzuki2013}. 
   The energy gap is formed, for $J =0$,  due to the symmetry-induced
  splitting between the $\Gamma_4^{ }$, $\Gamma_6^{(1)}$ and
  $\Gamma_5^{ }$, $\Gamma_6^{(2)}$ states, see Fig.\ \ref{Fig:LevelScheme_NpO2}.
   In the one-electron energy level scheme, the splitting between the $\Gamma_4^{}$ and $\Gamma_5$ singlet
  orbitals, which are degenerate under time-reversal symmetry,
  leads to the formation of the magnetic (111)-$T_{2g}^{-}$ octupole moment without
  inducing any magnetic dipole moment when the electrons occupy one of the singlet orbitals. This magnetic multipolar ordered
  state is expected to correspond to a singlet ground state in the
  \textit{L-S} coupling scheme under the (111)-$T_{2g}^{-}$ multipolar order,
  split from the paramagnetic $\Gamma_8$ quartet CEF ground
  state, see Fig.\ \ref{Fig:CEF_Multipole_NpO2}.
      The explicit calculations showed that, in the absence of any magnetic dipole moment, a
     magnetic $T_{2g}$ multipolar ordered state is obtained for NpO$_2$, which is
   sensitive to the Hund coupling parameter $J$ in the LDA+$U$ method. 
The Hund coupling induces the hybridization between the $\Gamma_{6}^{(1)}$ and
   $\Gamma_{6}^{(2)}$ $j=5/2$ one-electron $f$ orbitals, whose origins are the
  $\Gamma_7$ and $\Gamma_8$ paramagnetic $j=5/2$ $f$-orbitals as can be
     observed from the DOS for the one-electron orbitals calculated with
     $J = 0.5$\,eV in Fig.\ \ref{Fig:DOSNpO2_Order} (b).
    The Hund's coupling related hybridization of the $\Gamma_{6}^{(1)}$ and
   $\Gamma_{6}^{(2)}$  $f$ orbitals leads to a dependence of the multipolar moment on $J$, which is illustrated
   in Fig.\ \ref{Fig:Multipole_NpO2}.
   This coupling enhances the magnetic triakontadipole (rank-5
  multipole) moment and suppresses the magnetic octupole (rank-3
  multipole) moment (see Fig.\ \ref{Fig:Multipole_NpO2}(b)) in correspondence to the computed change in the occupations of the $\Gamma_{6}^{(1)}$ and
   $\Gamma_{6}^{(2)}$ orbitals, and the off-diagonal  ($\Gamma_{6}^{(1)}$,$\Gamma_{6}^{(2)}$) matrix elements
   (Fig.\ \ref{Fig:Multipole_NpO2}(a)). These findings show that, when
     the Hund's rule coupling $J$ is turned on, the magnetic 
      distribution is dominated by the rank-5
     triakontadipole, which becomes enhanced, whereas the rank-3
     octupole moment is suppressed. Hence, the $T_{2g}^{-}$
     triakontadipole moment can be regarded as the leading magnetic
     order parameter,  a finding that is consistent with the analysis of
     inelastic neutron scattering
     measurements~\cite{santini2006,magnani2008}. The computed nonzero
     electric quadrupole moment, that appeared as the secondary
     order parameter for the magnetic multipolar order (See the
     Appendix), is in accordance with 
  resonant x-ray scattering experiments~\cite{mannix1999,paixao2002, caciuffo2003}.

  The spacial distributions of the charge and magnetic moment on the actinide ion, calculated with
 the LDA+$U$ method ($U=4$ eV and $J=0.5$ eV),  are shown in Fig.\ \ref{Fig:ChgSpn_U_Np}
 for UO$_2$ and NpO$_2$. The dominating dipolar magnetic nature of the 3${\bm q}$ antiferromagnetic
 ground state of UO$_2$ can be clearly recognized. The magnetic distribution on Np, conversely, 
 is very different. This spacial distribution, seen along the [111] axis, reveals several closely spaced areas with opposite directions of the local magnetization that alternate strongly around the Np ion.
   This plot illustrates that the main contribution to the unique hidden order parameter of NpO$_2$ stems from high-rank magnetic multipoles, and moreover, provides a  first-principles confirmation\cite{mtsuzuki2010,mtsuzuki2013} of their important role in the emergence of this anomalously ordered phase.

\section{Multipole fluctuations}
\label{Sec:MPfluct}
 It is already well-understood that the coupling of atomic dipole moments via the Heisenberg-Dirac exchange or Anderson super-exchange interactions leads to the occurrence of long-range magnetic order in condensed matter systems.
Similarly, the exchange coupling of higher-order atomic multipole moments must be responsible for the formation of long-range multipolar order, even in the absence of any dipole moments,\cite{santini2009,mtsuzuki2010} and fluctuations of the multipole order lead to phase transitions.
In general, the divergence of a susceptibility characterizes a second-order phase transition. Possible phase transitions can be found by investigating the propagation vector dependence of the susceptibilities in the nonmagnetic state. Recently, first-principles calculations of susceptibilities within the random-phase approximation (RPA) have been applied to $f$-electron compounds.~\cite{ikeda2012,ikeda2014} 
In following sections, we explain the procedure and
outline the study of the superconducting pairing mechanism
for CeCu$_2$Si$_2$ as an application
of the first-principles
calculation of the multipole susceptibility. 
 
\subsection{First-principles tight-binding model}
\label{Sec:tightbinding}
An effective tight-binding model of electronic states is helpful to
analyze the orbital degree of freedom in strongly correlated
systems
and investigate the electronic structures in the context of topological phenomena. Downfolding by the Wannier function~\cite{mostofi2008,kunes2010} is a useful technique to obtain the effective hopping matrix in a real-space representation. Generally, the Fourier transformation of the thus-obtained hopping matrix provides the following hopping Hamiltonian in reciprocal space, 
\begin{eqnarray}
 H_{0}&=&\sum_{{\bm k}}\sum_{\tau\tau'\ell\ell'\gamma\gamma'}h_{{\bm k}\tau\ell\gamma,\tau'\ell'\gamma'}a^{\dagger}_{{\bm k}\tau\ell\gamma}a_{{\bm k}\tau'\ell'\gamma'}\ \nonumber\\
      &=&\sum_{{\bm k}}\sum_{b}\varepsilon_{{\bm k}b}a^{\dagger}_{{\bm k}b}a_{{\bm k}b},
\end{eqnarray}
 where $a,~a^{\dagger}$ are annihilation and creation operators, respectively, $\varepsilon_{{\bm k}b}$ the band energies, and $
h_{{\bm k}\tau\ell\gamma,\tau'\ell'\gamma'}$ are the hopping Hamiltonian matrix elements.
 The hopping Hamiltonian can be decomposed into three parts that read as follows:
\begin{eqnarray}
 H_{0} &=&\sum_{{\bm k}}\Biggl\{ \sum_{\gamma\gamma'}^{(f)}h^{(f)}_{{\bm k}\gamma\gamma'}
  f^{\dagger}_{{\bm k}\gamma}f_{{\bm k}\gamma'} \nonumber \\
  & & +
  \sum_{\tau\ell\gamma,\tau'\ell'\gamma'}^{(c)}h^{(c)}_{{\bm k}\tau\ell\gamma,\tau'\ell'\gamma'} c_{{\bm k}\tau\ell\gamma}^{\dagger}c_{{\bm k}\tau'\ell'\gamma'}\nonumber \\
& & + \Biggl(\sum_{\tau\ell\gamma}^{(c)}\sum_{\gamma'}^{(f)} V_{{\bm
  k}\tau\ell\gamma\gamma'}c^{\dagger}_{{\bm k}\tau\ell\gamma}f_{{\bm
  k}\gamma'}+h.c.\Biggr)\Biggr\}\ ,
\end{eqnarray}
 where the first term denotes the correlated $f$ electrons, whose atom and
 orbital are represented with the index $(f)$, the second term denotes
the non-interacting conduction electrons, represented by $(c)$, and the third represents the
hybridization between them.
 The Hubbard type on-site interaction can be included for the correlated
 $f$ electron space, as follows,
\begin{eqnarray}
H'^{(f)}&=&\frac{U}{2}\sum_{m}^{(f)}\sum_{\sigma}f^{\dagger}_{m\sigma}f^{\dagger}_{m\bar{\sigma}}f_{m\bar{\sigma}}f_{m\sigma} \nonumber\\
&+&\frac{U'}{2}\sum_{m\ne m'}^{(f)}\sum_{\sigma\sigma'}f^{\dagger}_{m\sigma}f^{\dagger}_{m'\sigma'}f_{m'\sigma'}f_{m\sigma} \nonumber\\
&+&\frac{J}{2}\sum_{m\ne m'}^{(f)}\sum_{\sigma\sigma'}f^{\dagger}_{m\sigma}f^{\dagger}_{m'\sigma'}f_{m\sigma'}f_{m'\sigma} \nonumber\\
&+&\frac{J'}{2}\sum_{m\ne m'}^{(f)}\sum_{\sigma} f^{\dagger}_{m\sigma}f^{\dagger}_{m\bar{\sigma}}f_{m'\bar{\sigma}}f_{m'\sigma}.
\end{eqnarray}
The obtained Hamiltonian is a kind of Anderson lattice model with a material's specific band structure. The electronic state can be analyzed with a variety of numerical techniques in strongly correlated systems, such as the DMFT method. Here we introduce the RPA susceptibilities obtained by the orthodox diagram expansion in terms of the Hubbard interaction defined above. They provide the first step to analyze the electronic state, and possible phase transitions into superconductivity and multipolar orderings. 

\subsection{Multipolar susceptibility}
The generalized multipole-multipole correlation function in the correlated $f$-orbital space of two multipoles $A^{(f)}$ and $B^{(f)}$ can be evaluated from
\begin{eqnarray}
\langle\langle A^{(f)},B^{(f)}\rangle\rangle=\sum_{\gamma_1\gamma_2\gamma_3\gamma_4}
 A^{(f)}_{\gamma_2\gamma_1}\chi^{{\rm RPA}(f)}_{\gamma_1\gamma_2,\gamma_3\gamma_4}B^{(f)}_{\gamma_3\gamma_4} ,
\label{Eq:MPSuscept}
\end{eqnarray} 
where $\gamma_i$ denotes one of the spin-orbital components and specific representations of multipoles $A^{(f)}$ and $B^{(f)}$ are given following the local (on-site) symmetry.
 $\chi^{{\rm RPA}(f)}$ is the RPA susceptibility defined as:
\begin{eqnarray}
&&\chi^{{\rm RPA}(f)}_{\gamma_{1}\gamma_{2},\gamma_{3}\gamma_{4}}({\bm
 q})=\chi^{0(f)}_{\gamma_{1}\gamma_{2},\gamma_{3}\gamma_{4}}({\bm q})+ ~~~~~~~~~~\nonumber \\
&&~~~~~~~~  \sum_{\gamma_5\gamma_6\gamma_7\gamma_8}\chi^{0(f)}_{\gamma_{1}\gamma_{2},\gamma_{5}\gamma_{6}}({\bm q})\Gamma^{0(f)}_{\gamma_{5}\gamma_{6},\gamma_{7}\gamma_{8}} \chi^{{\rm RPA}(f)}_{\gamma_{7}\gamma_{8},\gamma_{3}\gamma_{4}}({\bm q})\ ,
\end{eqnarray}
\begin{eqnarray}
\chi^{0(f)}_{\gamma_1\gamma_2,\gamma_3\gamma_4}= -T\sum_{{\bm k},n}
 G^{0(f)}_{\gamma_1\gamma_3}({\bf k}+{\bm
 q},i\omega_n)G^{0(f)}_{\gamma_4\gamma_2}({\bm k},i\omega_n)\ ,
\end{eqnarray}
where $\Gamma^{0(f)}$ is the four point vertex of Hubbard type interaction and $G^{0(f)}({\bm k},i\omega_n)$ is the non-interacting Green function with the momentum ${\bm k}$ and Matsubara frequency $\omega_n$,
\begin{eqnarray}
 G^{0(f)}_{\gamma\gamma'}({\bm
  k},i\omega_n)=\frac{\langle (f)\gamma\mid{\bm k},b\rangle\langle {\bm
  k}, b\mid (f)\gamma'\rangle}{i\omega_{n}-\varepsilon_{{\bm k}b}+\mu} .
\end{eqnarray}

\noindent
In multi-orbital systems, vertex corrections can be crucial by
hybridizing not only different 
wave vectors ${\bm Q}$ and frequencies $\omega$ but also multi-channels such
as electric and magnetic channels. The RPA susceptibility usually
enhances a magnetic channel too much, and this can be improved by including
the mode-mode coupling through the vertex
corrections~\cite{ikeda2012,ikeda2014}.

\subsection{Superconducting gap}
 Multipole fluctuations have also been applied to
investigate theoretically the interaction mediating the Cooper pair formation of
superconductivity in heavy electron system~\cite{ikeda2015}.
 The gap function $\Delta^{(f)}({\bm k})$ is obtained by solving the linearized BCS gap equation,
\begin{eqnarray}
 \lambda\Delta^{(f)}_{\gamma\gamma'}({\bm k}) &=&- \sum_{{\bm
  k}'}\sum_{\gamma_1\gamma_2\gamma_3\gamma_4}V^{(f)}_{\gamma\gamma_1,\gamma_2\gamma'}({\bm k}-{\bm k}') \times \nonumber \\
  & &G^{(f)}_{\gamma_1\gamma_3}({\bm
  k}')G^{(f)}_{\gamma_2\gamma_4}({\bm k}')^{*}\Delta^{(f)}_{\gamma_3\gamma_4} ,
  \label{eq:gap}
  \end{eqnarray}
with $V^{(f)}_{\gamma_1\gamma_2,\gamma_3\gamma_4}({\bm k}-{\bm k}')$ the pairing potential defined by the multipolar fluctuations,  given as
  \begin{eqnarray}
 \!\! \! V^{(f)}_{\gamma_{1}\gamma_{2},\gamma_{3}\gamma_{4}}({\bm
  q})& = & \Gamma^{0(f)}_{\gamma_{1}\gamma_{2},\gamma_{3}\gamma_{4}}+ \nonumber \\
  & & \!\! \!  \!\! \! \sum_{\gamma_{5}\gamma_{6}\gamma_{7}\gamma_{8}}\Gamma^{0(f)}_{\gamma_{1}\gamma_{2},\gamma_{5}\gamma_{6}}\chi^{(f)}_{\gamma_{5}\gamma_{6},\gamma_{7}\gamma_{8}}({\bm
  q})\Gamma^{0(f)}_{\gamma_{7}\gamma_{8},\gamma_{3}\gamma_{4}} .
\end{eqnarray}
 $\chi^{(f)}({\bm q})$ is $\chi^{0(f)}({\bm q})$ for the second-order
 perturbation and $\chi^{{\rm RPA}(f)}({\bm q})$ for the RPA.
The transition temperature $T_{c}$ is obtained as the temperature when the
maximum eigenvalue $\lambda$ is unity~\cite{ikeda2010}.

The first-principles calculation of multipole fluctuations was applied to investigate the pairing mechanism of superconductivity in the heavy Fermion compound CeCu$_2$Si$_2$~\cite{ikeda2015} and possible order parameters of hidden order phase of URu$_2$Si$_2$~\cite{ikeda2012}. In both cases, the first-principles calculations show a crucial contribution of high-rank magnetic multipoles to the multipole susceptibility.

\subsection{CeCu$_2$Si$_2$}
\begin{figure}[t]
	\includegraphics[width=0.99\linewidth]{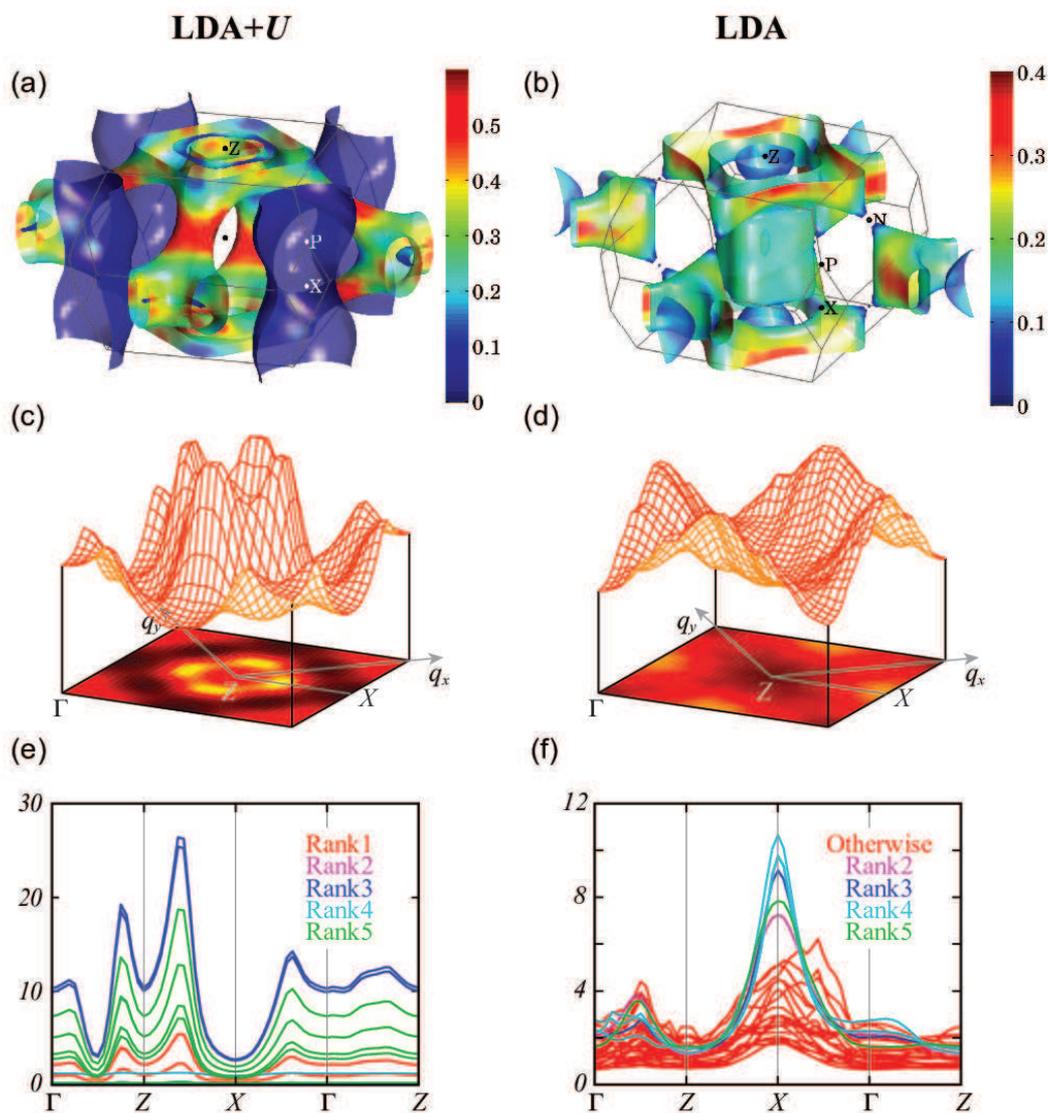}
	\caption{ (Color online) (Top) Calculated Fermi surface of nonmagnetic CeCu$_2$Si$_2$ colored by the Fermi velocity.
 (Middle) The in-plane magnetic dipole susceptibilities computed for
 ${\bm Q}=(q_x,q_y,0.5)$, and (Bottom) a complete set of multipole
 susceptibilities along the high-symmetry lines. The left-hand figures
 correspond to the LDA+$U$ case, and the right-hand panels to the ordinary LDA
 case. In (c), incommensurate peaks at around ${\bm Q}=(0.21,0.21,0.5)$ and
 the corresponding points are consistent with INS measurement (from
 Ref.~\cite{ikeda2015})
\textcopyright 2015 American Physical Society} 
	\label{CeCu2Si2}
\end{figure}
\begin{figure}[t]
	\includegraphics[width=\linewidth]{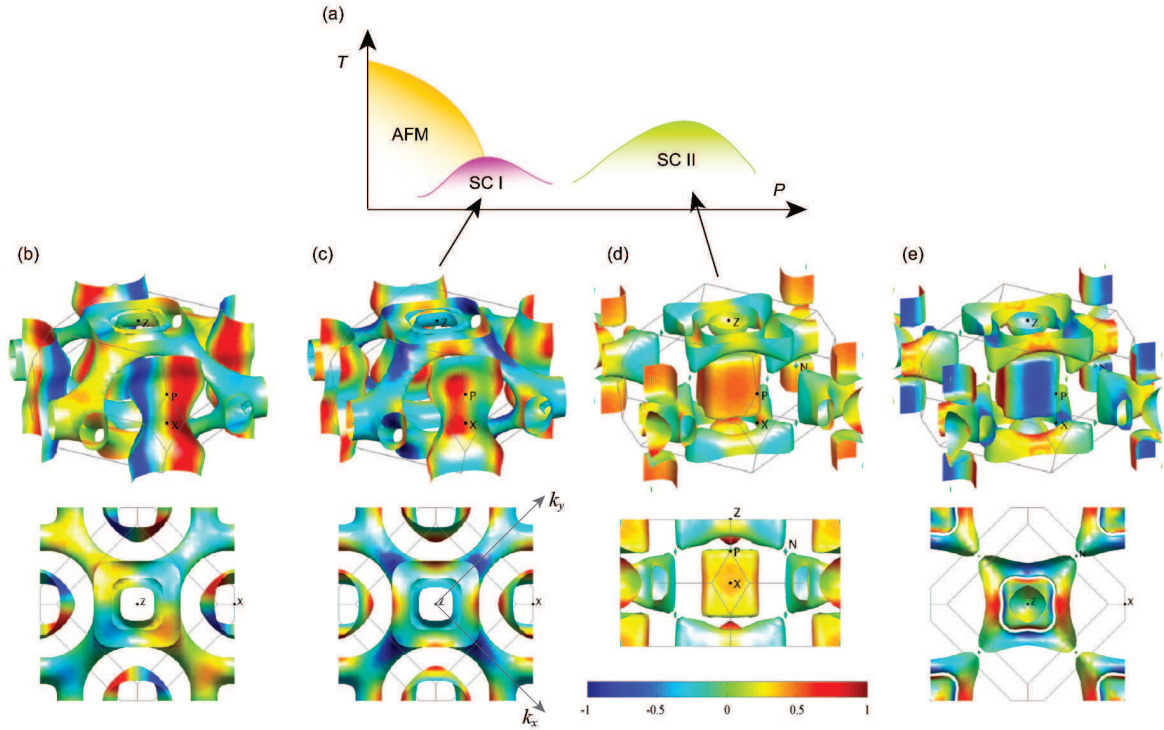}
	\caption{ (Color online) (a) Schematic pressure ($P$) -- temperature ($T$) phase diagram for 
 CeCu$_2$Si$_2$, showing the AFM phase and two superconducting phases. 
(b-e) Possible superconducting gap structures corresponding to order parameters of 
(b) $d_{x^2-y^2}$ $[\sim \cos (2k_x)-\cos (2k_y)]$,
(c) $s_\pm$ $[\sim \cos (2k_x)+\cos (2k_y)]$ from the LDA+$U$ Fermi surface, 
(d) another type $s_\pm$ $[\sim \cos (k_x)\cos (k_y)\cos (k_z)]$ in the LDA case,
and (e) $d_{xy}$-wave $[\sim \sin (k_x)\sin (k_y) \cos (k_z)]$ pairing
 states. (From Ref.~\cite{ikeda2015}).
\textcopyright 2015 American Physical Society}
	\label{CeCu2Si2gap}
\end{figure}
This compound is the first discovered heavy-Fermion superconductor~\cite{steglich1979}. This compound 
 crystallizes in the body-centered tetragonal  structure (space group
 No.\ 139, $I4/mmm$). The BCS-like specific heat jump at the transition
 temperature ($T_c \le 1 $ K) indicates that the correlated heavy
 electrons form the Cooper pairs. It has long been considered that this
 is a prototypical example of a nodal $d$-wave superconductor mediated
 by magnetic fluctuations, based on several experimental observations,
 such as the $T^2$ behavior of low-temperature specific
 heat~\cite{bredl1983,arndt2011}, no coherence peak just below $T_c$ and
 the $T^3$ behavior in NMR relaxation rate
 $1/T_1$~\cite{kitaoka1986,ishida1999,fujiwara2008}, and the
 resonance-like enhancement in inelastic neutron scattering (INS)
 measurements~\cite{stockert2011} and so on.\cite{pfleiderer2009}
 However, in contrast to these observations, the recent specific heat
 measurement~\cite{kittaka2014}, penetration depth, and thermal
 conductivity measured down to very low
 temperatures~\cite{yamashita2017} have indicated that the
 superconducting order parameter is fully gapped. Consequently, the gap
 structure and the pairing mechanism in this material have been hotly
 debated. Although generally the Fermi surface (FS) topology is
 crucially important for superconductivity, it is not yet so clear for
 CeCu$_2$Si$_2$. Theoretically, two or three types of FSs have been
 proposed based on the ordinary LDA calculations~\cite{harima1991},
 LDA$+U$ method~\cite{kittaka2014} and the renormalized band
 theory~\cite{zwicknagl1993} (See Figs.\ \ref{CeCu2Si2}(a) and
 (b)). Figures \ref{CeCu2Si2}(c) and (d) depict the corresponding
 magnetic dipole susceptibilities calculated within
 RPA~\cite{ikeda2015}. They show the characteristic $\bm Q$ structure
 due to Fermi surface nesting; an incommensurate peak at ${\bm
 Q}=(0.21,0.21,0.5)$ in Fig.\ \ref{CeCu2Si2}(c), and a hump at around
 the $X$ point in Fig.\ \ref{CeCu2Si2}(d). The former is consistent with
 the peak position observed in the INS measurement~\cite{stockert2011},
 while the latter has not been observed. This implies that the FS as
 obtained from the LDA$+U$, that shows similar trends with that
 of the renormalized band theory, is more appropriate rather than the
 LDA FS. 
Figures \ref{CeCu2Si2}(e) and (f) illustrate multipole susceptibilities along high-symmetry lines, computed with Eq.\ (17). 
The dominant fluctuations are of octupole type in Fig.\ \ref{CeCu2Si2}(e) and of hexadecapole type in Fig.\ \ref{CeCu2Si2}(f). Furthermore, from Eq.\ (\ref{eq:gap}), these high-rank multipole fluctuations can lead to the $d_{x^2-y^2}$-wave and $s_\pm$-wave pairing states (See Fig.\ \ref{CeCu2Si2gap}). The latter is compatible with the full-gap nature observed recently.~\cite{kittaka2014}
 It deserves to be mentioned, though, that quite recently, a study of the impurity effect by electron irradiation demonstrated the robustness of superconductivity against disorder~\cite{yamashita2017,takenaka2017}, indicative of conventional $s$-wave pairing without sign-change. Thus, this issue is still to be clarified along with the FS topology. 

\section{Hidden order}
\label{Sec:HiddenOrder}

\subsection{The hidden order phase of URu$_2$Si$_2$}

The mysterious hidden order (HO) phase of the heavy-Fermion URu$_2$Si$_2$ has drawn much attention in condensed matter physics during the last three decades.  URu$_2$Si$_2$, which crystallizes in the body-centered tetragonal  structure (space group No.\ 139, $I4/mmm$), exhibits a clear second-order phase transition 
 at $T_{\rm HO}$=17.5 K, which is signaled by a large entropy loss,~\cite{palstra1985, maple1986, schlabitz1986} 
 a large reduction of the carrier number,~\cite{palstra1985,behnia2005, kasahara2007}
 and the emergence of a partial gap.~\cite{escudero1994,rodrigo1997,schmidt2010,aynajian2010}
  Neutron and X-ray scattering investigations revealed that there exists
 a vanishingly small dipole magnetic moment in the HO phase
 \cite{broholm1987,isaacs1990,mason1990}, which depends on sample purity
 but is much too small to explain the large entropy change seen at the
 phase transition \cite{buyers1996}. 
Unravelling the origin of the HO phase of URu$_2$Si$_2$ subsequently became the object of intense investigations, ongoing as much today. The underlying origin of the low-temperature order could not unambiguously be disclosed, even after thirty years (see Refs.\ \cite{mydosh2011,mydosh2014} for recent reviews on the status of the HO). 

  NpO$_2$ has been the archetypal example of a compound displaying a HO phase, whose
 mystery has been solved in recent years, as discussed in Sec.\
 \ref{Sec:IntroAnO2}.
   Applying the lessons learned from the great efforts paid to identify the HO phase of NpO$_2$
    to solve the origin of the enigmatic HO phase of URu$_2$Si$_2$ can be helpful,
 however, it is important to realize that some properties of the electronic and ordered states in
 URu$_2$Si$_2$ are different from those of NpO$_2$ which makes the situation for the identification of
  the order parameter more complex.
  One of the clear differences between the hidden order phase of
 URu$_2$Si$_2$ and the low-temperature magnetic multipolar phase of
 NpO$_2$ is the strong itinerant character of the $f$-electrons in
 URu$_2$Si$_2$ (see, e.g., Refs.\ \cite{oppeneer2010,fujimori2016}),
 as compared to the localized $5f$ character of NpO$_2$ (see, e.g.\ Refs.\ \cite{santini2009, caciuffo2003, sakai2005}),
 which causes an ambiguity for the experimental determination of the
 applicable CEF level scheme. 
  Thus far, no evidence of localized $f$ states  or CEF excitations in URu$_2$Si$_2$
 could be detected.\cite{mydosh2011}
  Another crucial difference between the two compounds is the symmetry
 property of the ordinary antiferromagnetic order in URu$_2$Si$_2$\cite{mtsuzuki2014}. 
 The latter property is in the focus of the Appendix.

\begin{figure}[t]
\begin{center}
	\includegraphics[width=0.6\linewidth]{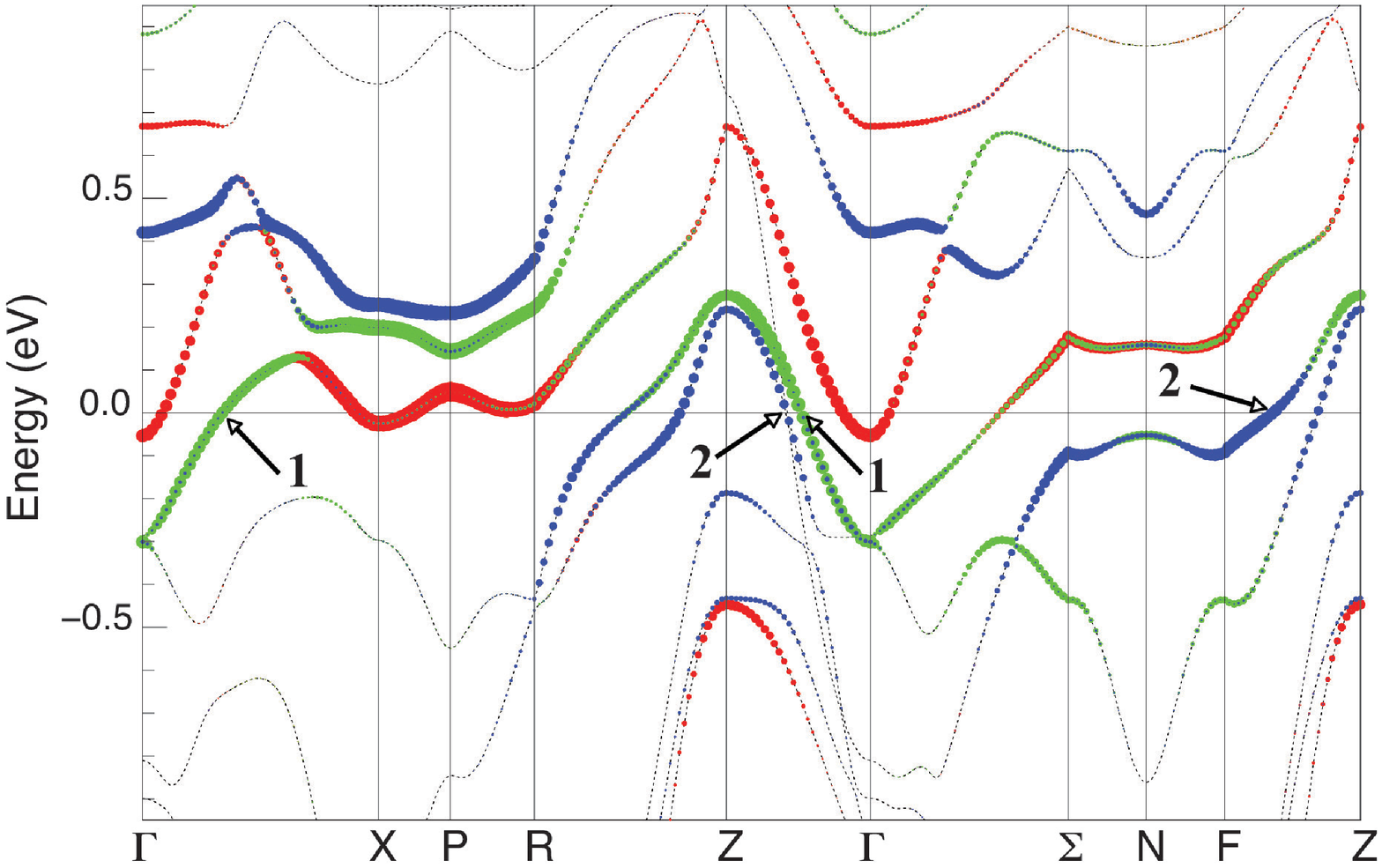}
	\caption{ (Color online) Band structure of nonmagnetic URu$_2$Si$_2$ in the BCT structure. Colors of the fat bands denote the uranium 
	$5f_{5/2}$ character; specifically, green depicts the $j_z=\pm 5/2$ character, blue the $j_z=\pm 3/2$  character, and 
	red the $j_z=\pm 1/2$ character. The indices 1 and 2 denote the nested bands at the Fermi energy that are predominately formed 
	from $j_z = \pm 5/2$ and $ \pm 3/2$  states (from Ref.\
 {\cite{oppeneer2011}}).
\textcopyright 2011 American Physical Society}
	\label{URu2Si2band}
\end{center}
\vspace{20pt}
\begin{center}
	\includegraphics[width=0.8\linewidth]{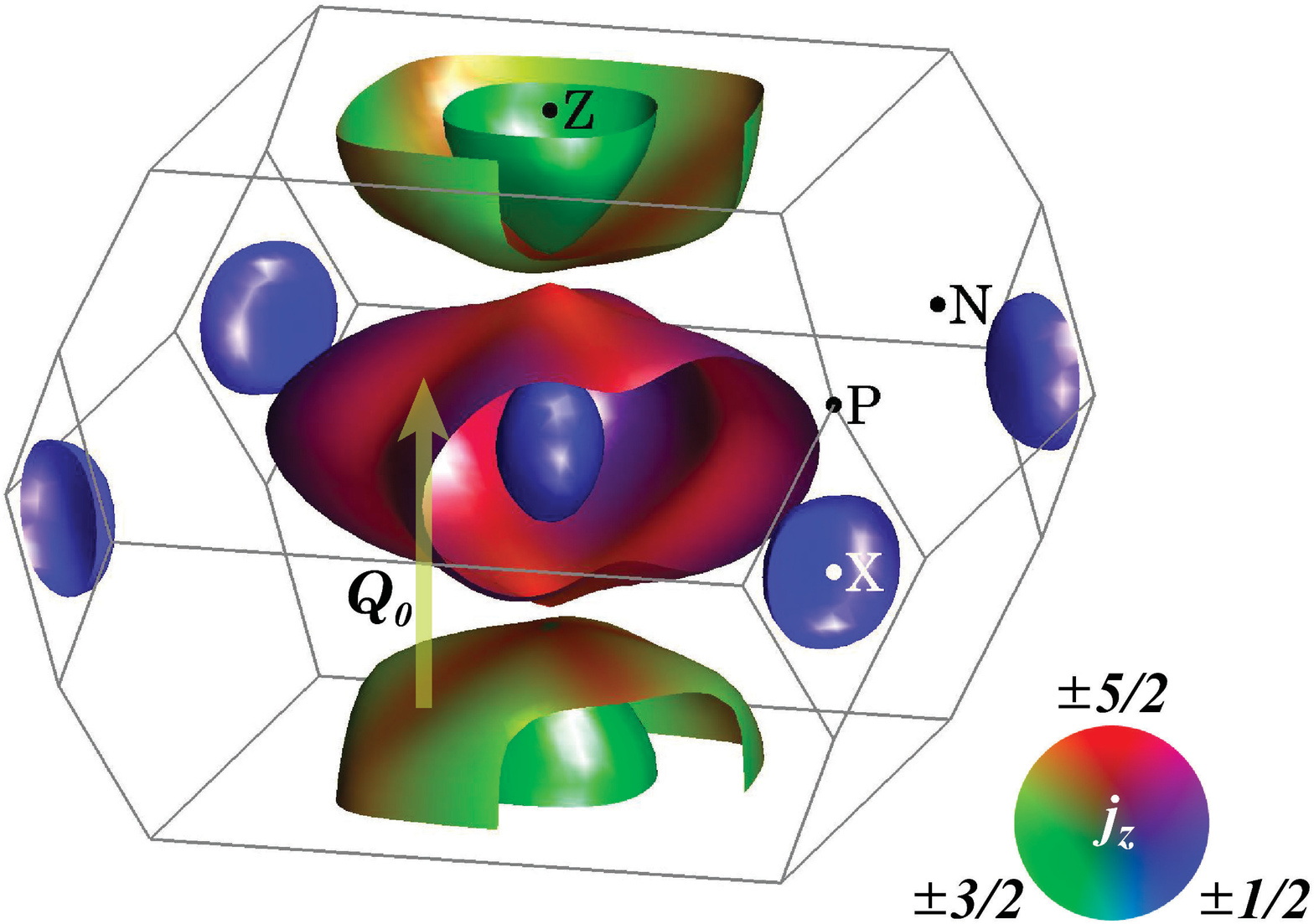}
	\caption{ (Color online) Calculated (LDA) Fermi surface of URu$_2$Si$_2$ with
 the amount $j$=5/2 orbital components indicated by the colors. (From
 Ref.~\cite{ikeda2012}).
\textcopyright 2012 Macmillan Publishers Limited.}
	\label{URu2Si2fs}
\end{center}
\end{figure}

\begin{figure}[t]
	\includegraphics[width=1.\linewidth]{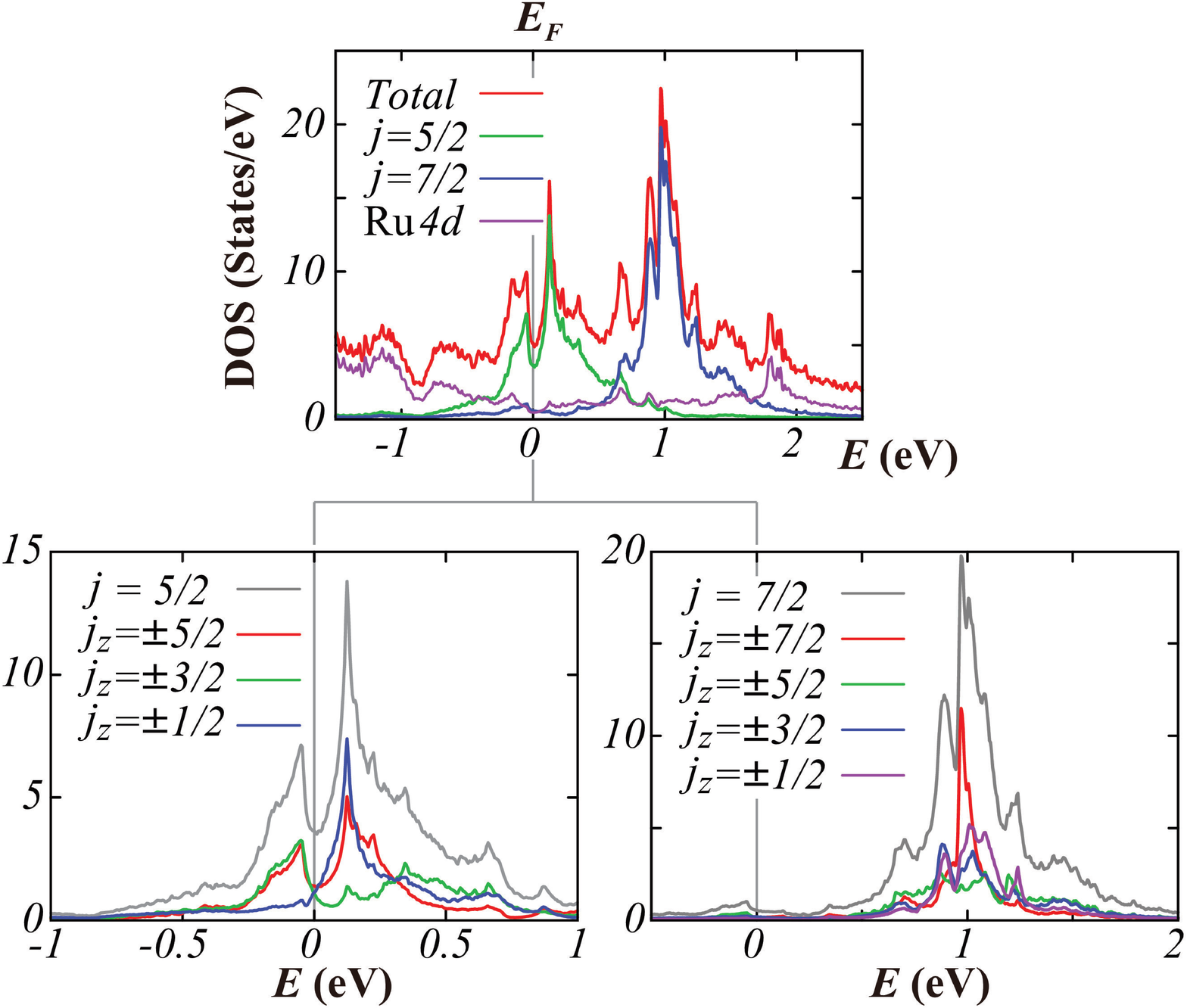}
	\caption{ (Color online) Partial density of states of URu$_2$Si$_2$ in the
 nonmagnetic state. The lower panels show the $j_z$-projected partial
 density of states for the $j=5/2$ and $j=7/2$ manifolds (from Ref.\
 \cite{ikeda2012}).
\textcopyright 2012 Macmillan Publishers Limited.}
\label{Fig:DOS}
\end{figure}

The understanding of the electronic properties of  URu$_2$Si$_2$ has nonetheless increased much. Among the interesting characteristics of  URu$_2$Si$_2$ are the proximity of an antiferromagnetic (AFM) ordered phase that can be induced by a small pressure (of $\sim$0.6 GPa)
\cite{amitsuka1999,matsuda2001,amitsuka2003,motoyama2003,amato2004,amitsuka2007, takagi2007,motoyama2008,hassinger2008,butch2010}. Interestingly, the Fermi surfaces of URu$_2$Si$_2$ hardly change between the HO and AFM phases as obtained from de Haas-van Alphen and Shubnikov-de Haas measurements.~\cite{ohkuni1999,nakashima2003,jo2007,hassinger2010} In spite of the nearly identical Fermi surfaces
an unconventional superconducting phase only emerges out of the HO phase at temperatures below 1.5 K.~\cite{kasahara2007,yano2008} A further similarity of the phases is the appearance of a longitudinal spin wave mode at wavevector ${\bm Q}_0 = (0, 0, 1)$ in the HO phase which freezes to become the static long-range  order of the AFM phase \cite{villaume2008,bourdarot2010}.
One of the crucial questions in the quest for the origin of the HO has
been: which symmetry is spontaneously broken at the HO transition? 
A number of recent measurements have provided evidence for a breaking of the body-centered translation vector in the HO phase, leading to a doubling of the unit cell and a folding of the Brillouin zone
\cite{yoshida2012,meng2013,buhot2014}. 
  The breaking of time-reversal symmetry could
 not unambiguously be established (see, e.g. Refs.\
 {\cite{walker1993,takagi2012,schemm2015}}); the spuriously small
 magnetic dipole moment in the HO phase that could be of extrinsic
 origin prohibits an unambiguous identification.
 A remarkable set of experiments led to the conclusion that the four-fold symmetry of the tetragonal basal plane is broken and that the HO is hence a ``nematic'' phase~\cite{okazaki2011,tonegawa2012,kambe2013,tonegawa2014,riggs2015}. 
Although this symmetry-breaking was reported e.g.\ in X-ray Bragg scattering\cite{tonegawa2014} and NMR\cite{kambe2013} experiments, the results of other microscopic probes such as the NQR spectra~\cite{saitoh2005,mito2013} still remain controversial.
Furthermore, recent electronic Raman experiments~\cite{buhot2014,kung2015,kung2016} revealed a sharp low-lying excitation with $A_{2g}$ symmetry. The static $A_{2g}$ Raman signal closely resembles the uniform $J_z$ susceptibility. In the HO phase, $A_{1g}$ mode is also active.  These findings may impose a strong constraint on the theoretical models.

So far, many theories have been proposed to explain the enigmatic HO phase of URu$_2$Si$_2$ (see Refs.\ \cite{mydosh2011,mydosh2014}). Among the proposed explanations, the possible formation of multipolar order on the uranium ion takes a prominent place~\cite{nieuwenhuys1987,santini1994,ohkawa1999,hanzawa2005,kiss2005,fazekas2005,hanzawa2007,cricchio2009,haule2009,haule2010,harima2010,miyake2010,thalmeier2011,kusunose2011,rau2012,ikeda2012,kung2015}. Other competing theories are based on the presence of a Fermi surface 
instability and a concomitant gap opening due to a rearrangement of itinerant $5f$ electrons near the Fermi energy~\cite{luthi1993,ikeda1998,chandra2002,mineev2005,varma2006,elgazzar2009,oppeneer2010,dubi2011,pepin2011,fujimoto2011,riseborough2012,das2012,hsu2013,kotetes2014}.
 From a theoretical view point, the dual nature of $f$ electrons, which can be itinerant or localized, makes the situation complicated. In addition,
experimental techniques cannot straightforwardly provide evidence for one particular model. 
 At present, as the primary HO parameter, dipolar magnetic
order $J_{z}$ can be excluded due to the presence of a bicritical point,
at $p<0.6$\,GPa where the HO temperature and N{\'e}el temperature meet the first-order phase
boundary of the HO-AFM transition~\cite{motoyama2003,amato2004,motoyama2008,hassinger2008,butch2010}, from which follows the different
symmetries for the order parameters of the HO and AFM phases due to 
lacking of a bilinear term of these order parameters in the Landau free energy~\cite{kiss2005, mineev2005}. 
Resonant X-ray scattering experiments could not observe any quadrupole~\cite{nagao2005,walker2011a}. 
So far, the existence of higher order electric/magnetic multipole could not be unambiguously detected\cite{khalyavin2014,wang2017}. 
 Under such situation, \textit{ab initio} calculations and the group theoretical approach~\cite{harima2010, mtsuzuki2014} can provide useful knowledge.

To predict the most likely multipoles,  \textit{ab initio} electronic structure calculations of URu$_2$Si$_2$ in the nonmagnetic and AFM phases have been performed.\cite{yamagami2000,elgazzar2009,oppeneer2010,oppeneer2011,ikeda2012,werwinski2014}
These calculations notably correctly reproduce many of the known
properties of URu$_2$Si$_2$, as e.g.\ details of the AFM phase, lattice
properties and Fermi surface. First-principles calculation based on the LDA
 well reproduced the de Haas-van Alphen and Shubnikov-de
 Haas experiments.\cite{elgazzar2009,oppeneer2010,hassinger2010}
The calculated LDA band dispersions of
URu$_2$Si$_2$ in the nonmagnetic phase are shown in Fig.\
\ref{URu2Si2band}. The colors highlight which bands have dominantly
uranium $j=5/2$ character \cite{oppeneer2011,ikeda2012}. 
 It can be recognized that the bands at the Fermi surface consist primarily of
$j_z= \pm 5/2, ~\pm 3/2$, and $\pm 1/2$ states. The $j_z=\pm 5/2$ and
$\pm 3/2$ bands (labeled 1 and 2) form two nested Fermi surface sheets
that are connected by the nesting vector ${\bm Q}_0 = (0,0,1)$ as shown
in Fig.\ \ref{URu2Si2fs}.
 It can be recognized that the $\Gamma$ and X-point pocket consist
 mainly of $j_z =\pm 1/2$ character, whereas the Z-point centered sheets
 contain dominantly $j_z= \pm 3/2$ character, and the large $\Gamma$
 -centered sheet $j_z = \pm 5/2$ character.

\begin{figure}[t]
\begin{center} 
	\includegraphics[width=0.6\linewidth]{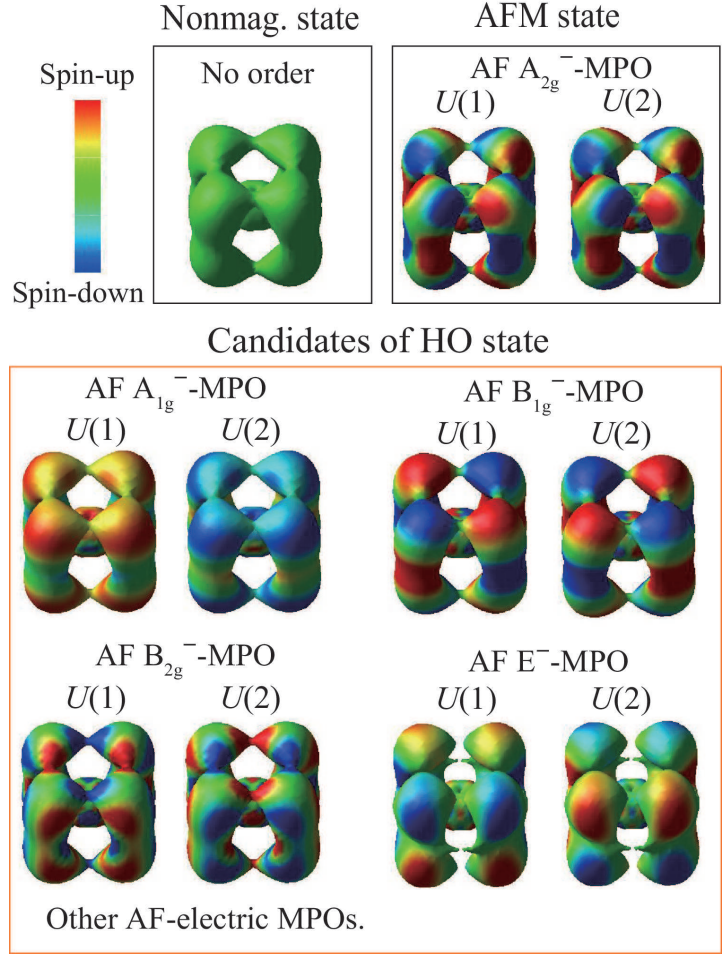}
	\caption{
 (Color online) Charge density and magnetization distribution of magnetic multipolar
 ordered (MPO) states in URu$_2$Si$_2$, computed with
 $U=2$\,eV and $J=0$ in the LDA+$U$ method. The magnetization
 distributions depict the spin-component along the [001] axis and are
 shown on the isosurfaces of the charge densities on the two uranium
 ions related to one another by the body-center translation. The
 distributions are viewed from the [111] direction. }
	\label{Fig:URu2Si2MP}
\end{center}
\end{figure}


In Fig.\ \ref{Fig:DOS} we show the calculated LDA density of states (DOS) in
an energy interval around the Fermi energy~\cite{ikeda2012}.  It can
again be observed that $5f$ states near the Fermi energy stem fully from
the $j=5/2$ sextuplet. 
 The Fermi level is in the hybridization gap. The $j=7/2$ multiplet is
 located at around 1 eV higher due to the S-O interaction. The $5f$
 occupation number is 2.07 for $j=5/2$ and 0.64 for
 $j=7/2$.~\cite{ikeda2012}
  The computed $5f$ occupation is consistent with several recent spectroscopic experiments.\cite{jeffries2010,booth2016}

 In Fig.\ \ref{Fig:URu2Si2MP} we show the magnetization distributions
 of possible AF magnetic multipolar ordered (MPO) states around each uranium atomic
 site as calculated by the LDA+$U$ method, assuming a
   Coulomb $U=2$ eV, which is relatively larger than theoretically
 expected values (less than 1.0\,eV~\cite{cricchio2009,oppeneer2010}) to
 emphasize the characters of the distributions.
   In this figure, the spacial magnetization distributions reflect the magnetic
 symmetries for each IREP. The charge distributions show
 no qualitative change for the $A_{1g}^{-}$, $A_{2g}^{-}$, $B_{1g}^{-}$,
 $B_{2g}^{-}$-MPO states by preserving the same symmetry with that of
 the nonmagnetic state, but it is deformed in the $E_{g}^{-}$ -MPO state
 reflecting the secondary induced ferroic $E_{g}^{+}$ electric multipolar order
 as discussed in the Appendix.
 
 As shown in the following section, the multipole RPA susceptibility (Fig.\ \ref{Fig:URu2Si2MPsuscept}) in this material indicated the dominant Ising-type $J_z$ correlation at commensurate ${\bm Q}_0 = (0,0,1)$ due to the Fermi surface nesting. In addition, some high-rank multipole correlations are enhanced due to the $j_z$ components on the nested Fermi surface, indicative of the $E_{g}^-$ triakontadipole as the first candidate and $A_{1g}^-$ triakontadipole as the second candidate for the possible multipole. 
Interestingly, the antiferroic configuration of the $E_{g}^-$ and $A_{1g}^-$ multipoles 
have completely different character from the view point of
symmetry, as discussed in detail in the Appendix. 
First, the $E_{g}^-$ multipole moments contain both the magnetic dipole and
octupole moments, but the lowest rank multipole moment which belongs to
the $A_{1g}^-$ IREP contains only the triakontadipole moment. 
This means that ordering of $E_{g}^{-}$ triakontadipole moments can bring with it these
lower rank multipoles, though the rank-1 magnetic dipole moments are expected to
be very small on the basis of the LDA+$U$
calculations~\cite{mtsuzuki2014}, while the $A_{1}^{-}$ multipole
moments cannot.
Second, since the multipoles of $E_{g}^{-}$ IREP are
transformed as an axial vector orthogonal to the four-fold rotation axis,
this ordering breaks the four fold symmetry of the system while the
multipoles of $A_{1g}^{-}$ representation
 invert their sign only through the time-reversal operation, and hence the corresponding 
order preserves the local point group symmetry.

\subsection{Multipole susceptibilities}
\begin{figure}[t]
\begin{center}
	\includegraphics[width=0.8\linewidth]{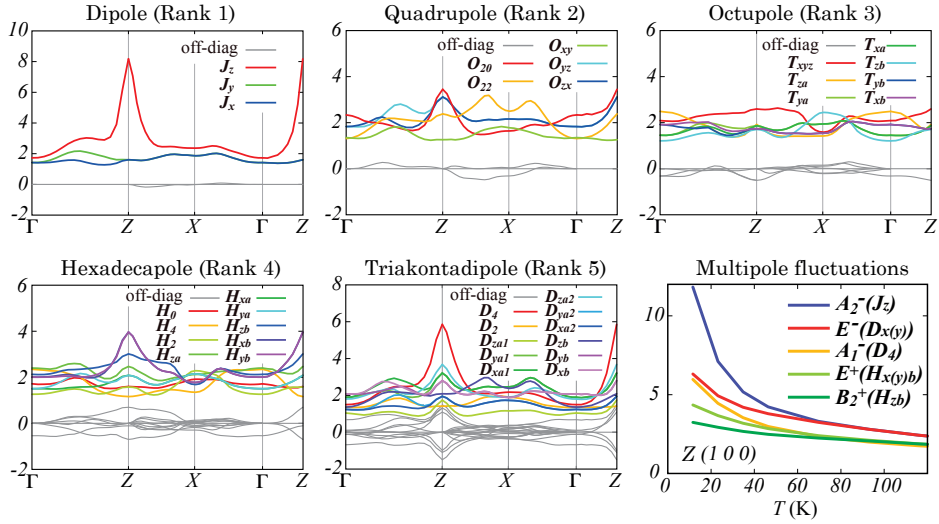}
	\caption{ (Color online) Multipole susceptibilities computed for URu$_2$Si$_2$
	 along high-symmetry lines in the BZ, using Eq.\ (\ref{Eq:MPSuscept}), and computed
	 temperature dependence of the most divergent susceptibilities for each IREP at the $Z$ point. 
	 The on-site multipoles $J$, $O$, $T$, $H$, and $D$ denote
 the magnetic dipole (rank-1),  electric quadrupole (rank-2), magnetic octupole (rank-3), electric
 hexadecapole (rank-4), and magnetic triakontadipole (rank-5),
 respectively, see Ref.\ \cite{ikeda2012} for details of the specific representations.
Note that in a generic $k$ point, all rank multipoles with identical
 IREP in the little group of $k$ can be mixed. 
It is implied by the presence of off-diagonal correlations.
}
	\label{Fig:URu2Si2MPsuscept}
\end{center}
\end{figure}
On the basis of the itinerant band structure and adopting a Coulomb $U \sim 1$ eV, for
the nonmagnetic state of URu$_2$Si$_2$,
multipole fluctuations have been studied within the RPA and
beyond~\cite{ikeda2012}. 
Figure \ref{Fig:URu2Si2MPsuscept} shows calculated multipole susceptibilities along
the high symmetry lines. In the dipole susceptibilities, we can see that
the $J_z$ susceptibility shows a peak structure at the $Z$ point and a
hump structure at the incommensurate wavevector ${\bm Q}_1=(1.4,0,0)$,
while the in-plane $J_{x(y)}$ susceptibilities are inactive. These
specific $\bm Q$ vectors and the remarkable Ising-type anisotropy are
consistent with the INS measurements~\cite{wiebe2007,villaume2008}. In
addition, some high-rank multipole susceptibilities, such as that of
hexadecapole and triakontadipole, are also enhanced at the $Z$
point. These characteristic features come from the FS nesting and
$j=5/2$ $f$-orbital characters constructing the FS. Among these enhanced
susceptibilities,
the most divergent susceptibility provides a candidate for the HO
parameter. Generally, since susceptibilities with the same IREP are
mixed each other, the most divergent susceptibility is evaluated by
diagonalizing $\hat\chi({\bm{Q}}_0)$. As a consequence, antiferro $E_{g}^{-}$
and $A_{1g}^{-}$ multipolar ordered states with mainly
triakontadipole components are obtained as candidates for the HO
state.
These order parameters are consistent with some observations that might
imply time-reversal symmetry breaking, such as the internal field in
NMR~\cite{takagi2012}, $\mu$SR~\cite{kawasaki2014}, and the polar Kerr
effect~\cite{schemm2015}, though it is not yet definitely established
whether these experimental observations are intrinsic in the HO phase or not.
Moreover, the $E_{g}^{-}$ state is consistent with the reported nematic feature~\cite{okazaki2011,tonegawa2012,kambe2013,tonegawa2014,riggs2015}, but induces in-plane dipole $J_{x(y)}$ moments, which are of the order of $10^{-4}\mu_B$, experimentally.~\cite{takagi2012,das2013,metoki2013,ross2014}

\begin{figure}[t]
\begin{center}
	\includegraphics[width=0.8\linewidth]{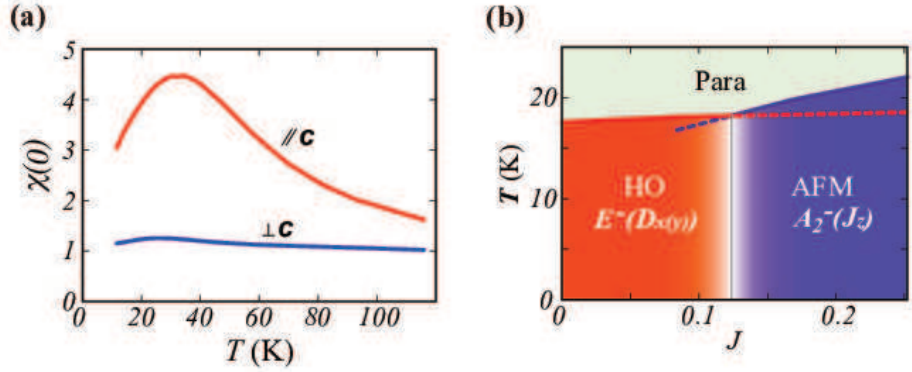}
	\caption{ (Color online) (a) Temperature dependence of uniform
 susceptibility obtained from beyond RPA calculations. (b) The phase
 diagram for the $E^-$ order parameter as a function of the Hund's
 coupling parameter $J$ (from Ref.\ \cite{ikeda2012}).
\textcopyright 2012 Macmillan Publishers Limited.} 
	\label{URu2Si2}
\end{center}
\end{figure}
Beyond the RPA calculations, including vertex corrections like the Maki-Thompson and the Aslamazov-Larkin terms, we can obtain the Ising-type uniform susceptibility (Fig.\ \ref{URu2Si2}(a)), and near degeneracy of the $E_{g}^-$ HO state and $A_{2g}^-$ AFM state (Fig.\ \ref{URu2Si2}(b)).~\cite{ikeda2012} This is the first example that the large Ising anisotropy in the static temperature-dependent magnetic susceptibility~\cite{palstra1985} was discussed on the basis of the first-principles electronic structure. 
 The phase diagram of near-degenerate $E_g^-$ and $A_{2g}^-$ orderings is very similar to the pressure-temperature phase diagram~\cite{motoyama2003,amato2004,motoyama2008,hassinger2008,butch2010}. This may indicate that the $E_g^-$ multipole is the HO order parameter.
 On the other hand, in the LDA+DMFT approach~\cite{haule2009}, $A_{2g}^+$
 hexadecapole order was discussed as the HO parameter. 
However, this multipole is not obtained from electronic structure calculations based on the (nearly) itinerant $f$-electron picture. 
The discrepancy is an issue to needs to be solved in the future. 
 In this regard, recent analyses of a possible CEF level scheme in nonresonant inelastic X-ray scattering~\cite{sundermann2016} and  in elastic constant measurements~\cite{kuwahara1997,yanagisawa2013} could be helpful to understand the formation of the heavy-fermion state.

\section{Cluster multipole theory}
\label{Sec:ClusterMP}
\subsection{Cluster multipole and physical phenomena of magnetic
  compounds}
\label{Sec:DefCMP}

   Macroscopic physical phenomena are customary discussed with respect their specific order
  parameters; of recent interest in condensed matter physics are composite magnetic monopole moments and 
  coupled electric and magnetic order parameters (e.g., in multiferroics)~\cite{shindou2001,ederer2007,spaldin2008,castelnovo2008,khomskii2012,spaldin2013,hayami2014,hayami2016,sumita2017,watanabe2017}.

  A new area, in which multipole moments recently gained importance, is
   that of the Hall conductivity in antiferromagnets.
   The anomalous Hall effect (AHE) is usually observed in ferromagnetic
  metals, and therefore it has traditionally been believed to be related to the nonzero ordered dipole
  magnetization. Meanwhile, the AHE has recently
  been studied for a certain type of noncollinear antiferromagnetic
materials~\cite{nagaosa2010, yoshii2000, taguchi2003_1, yasui2006,
  tomizawa2009, tomizawa2010, taguchi2003_2, taguchi2004, chen2014, kubler2014, nakatsuji2015, kiyohara2016, nayak2016}.
  In particular, a large anomalous Hall conductivity (AHC) has been observed for 
  noncollinear magnetic phases with only little resulting dipolar
   magnetization
  due to their noncollinear antiferromagnetic arrangement. These
   compounds have a vanishingly small macroscopic in-plane
   magnetization, of about M$\sim$0.002 $\mu_{B}$ for
   Mn$_3$Sn~\cite{tomiyoshi1982,brown1990} and M$\sim$0.005 $\mu_{B}$ for Mn$_3$Ge~\cite{yamada1988,nayak2016,kiyohara2016}.
   Previously, non-co-planer magnetic alignments characterized by a finite scalar
  spin chirality was known to induce the AHE~\cite{shindou2001}, but the co-planer magnetic
  configuration of Mn$_3$$Z$ has no such scalar spin chirality, which raises the question where the large AHC originates from.
   The magnetic degree of freedom which induces the AHE in
  Mn$_3$$Z$ was recently identified by introducing the concept of the cluster
  multipole moment, which is an extension of the ordinary magnetic
  multipole moment that characterizes the magnetization distribution around
  an atomic site~\cite{mtsuzuki2017}.

\begin{figure}[t]
	\includegraphics[width=1.0\linewidth]{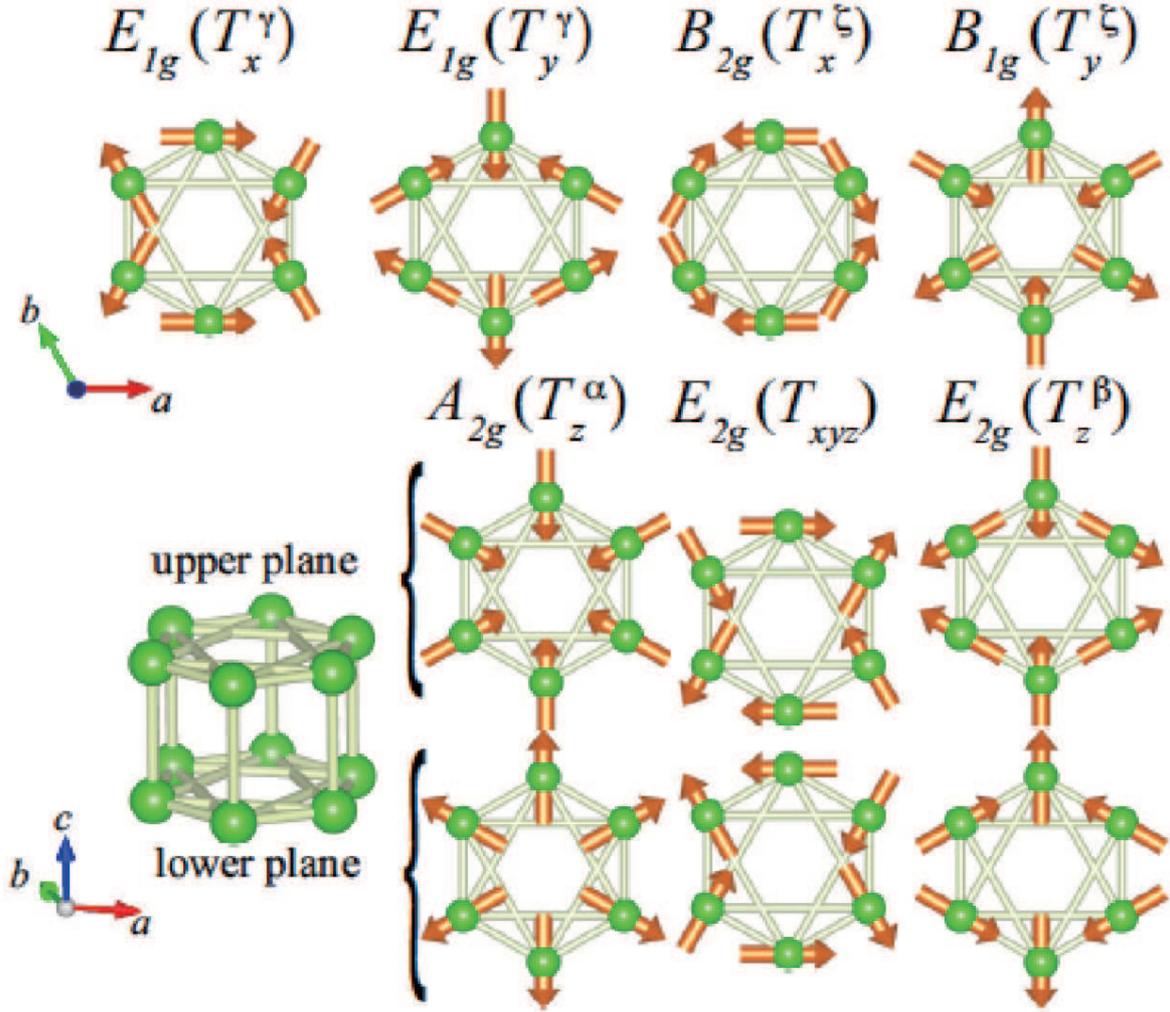}
	\caption{ (Color online) Examples
 of in-plane magnetic configurations of the hexagonal crystal structure
 that have zero cluster magnetic dipole moment but nonzero  magnetic
 octupole moments, classified according to the $D_{6h}$
 IREPs (from Ref.\ \cite{mtsuzuki2017}).
\textcopyright 2017 American Physical Society} 
	\label{Fig:CMP_Hex}
\end{figure}
   Cluster multipole theory offers an elegant way to obtain the magnetic order parameters
   that classify the broken magnetic point group symmetry due to
   the presence of the specific magnetic configuration in the crystal.
   To calculate these, one first has to determine the proper atomic
   clusters on which the cluster multipole moments are defined.
    For the magnetic order characterized by the ordering vector ${\bm
   q}={\bm 0}$, an atomic cluster in a crystal is defined as the atoms
   transformed in each other by the point symmetry operations
   in the space group ${\mathcal G}$ of the crystal system.
    In case of ${\bm q}\ne {\bm 0}$ order, whose magnetic unit cell is
   larger than the crystal unit cell, ${\mathcal G}$ is replaced, in
   the same context, with the
   subgroup of the crystal space group reflecting the reduced
   translation and rotation symmetry operations with the super cell.
  The operation elements of the space group ${\mathcal G}$ contains the
  translational operations combined with the point group
  operations. Symmorphic space groups have only the primitive translation operations which characterize the
  translation period. 
  In this case, the atomic clusters consist of the atoms that are related to 
  one another by the point group operation of ${\mathcal G}$ in the
  magnetic unit cell.  The other space groups ${\mathcal G}$,
  nonsymmorphic space groups, contain
  translation operations which differ from the primitive translations
  by a successive point group operation.
  These space groups can be decomposed as
\begin{eqnarray}
 {\mathcal G} = \sum_{i=1}^{N_{\rm coset}}\{R_{i}|{\boldsymbol \tau}_{i}\}{\mathcal H}\ ,
\label{Eq:coset} 
\end{eqnarray}
 where ${\mathcal H}$ is the maximal symmorphic subgroup of ${\mathcal
 G}$. An atomic cluster is defined as the atoms related to one another by
 the point group operations of ${\mathcal H}$ in the magnetic unit cell. 
 Note that the origin of the cluster is determined, except the degree of freedom to choose the origin satisfying the space group operation, as the spacial point
  which satisfy all of the point group symmetry of ${\mathcal H}$. 
   Other clusters in the crystal are defined as well after
 shifting the origin by the translation ${\boldsymbol \tau}_{i}$ in Eq.\
 (\ref{Eq:coset}) and/or by the primitive translation operations.

   Once the clusters have been identified the magnetic configuration in each
   cluster can be characterized by cluster multipole moments~\cite{mtsuzuki2017}, analogous
   to the local magnetic multipole moments of magnetic distributions on an atom, that are defined as 
\begin{eqnarray}
      M_{pq}^{(\mu)}\equiv
       \sqrt{\frac{4\pi}{2p+1}}\sum_{i=1}^{N^{(\mu)}_{\rm atom}}\nabla_{i} (|{\bm
       R}_{i}|^{p}Y_{pq}(\theta_i,\phi_i)^{*}) \cdot {\bm
       m}_{i}\ ,
\label{Eq:CMP}
\end{eqnarray}
 where $N^{(\mu)}_{\rm atom}$ is the number of atoms of the $\mu$-th
cluster, ${\bm m}_{i}$ is a magnetic moment on the $i$-th
atom, $\nabla_{i}\equiv\frac{\partial}{\partial {\bm R}_{i}}$, ${\bm
R}_i\equiv(X_i,Y_i,Z_i)$ is the position of the $i$-th atom, $Y_{pq}$
are the spherical harmonics, and $R_{i}$, $\theta_i$ and $\phi_i$ are the
distance, polar angle and azimuthal angle, respectively, of
the $i$-th atom.
  Figure \ref{Fig:CMP_Hex} shows examples of in-plane magnetic configurations
 characterized by cluster octupole moments classified according to the
 $D_{6h}$ IREPs~\cite{mtsuzuki2017}.
 Summing up the multipole moments in all the clusters, one obtains the
 order parameters reflecting the broken magnetic point group symmetry
 caused by the presence of the periodic magnetic configuration in the
 crystal.
  The macroscopic magnetization of the cluster multipole moments is
  thus obtained by summing the cluster multipole moments for all atomic
 clusters,
\begin{eqnarray}
      M_{pq}=\frac{N^{\rm u}_{\rm atom}}{N^{\rm c}_{\rm atom}}\frac{1}{V}\sum_{\mu=1}^{N_{\rm cluster}}M_{pq}^{(\mu)}\ ,
\label{Eq:CMPsum}
\end{eqnarray}
    where $N_{\rm atom}^{\rm u}$ is the number of atoms in the magnetic
  unit cell, $N_{\rm cluster}$ is the number of inequivalent clusters, which
  is the multiplication of the number of symmetrically inequivalent atoms by $N_{\rm coset}$, and $N_{\rm atom}^{\rm c}$ is the number of atoms contained in these clusters.
  The macroscopic magnetization of the cluster multipoles 
  is classified according to the point group symmetry of the crystal in
  the same way as the local multipole moments on atoms~\cite{shiina1997,kusunose2008}.
  Since the appearance of macroscopic physical phenomena generated by
  magnetic states is determined by the magnetic point group symmetry, these
  physical phenomena could be characterized by the macroscopic
  magnetization of the cluster multipole moments.

\begin{figure}[t]
 	\includegraphics[width=1.0\linewidth]{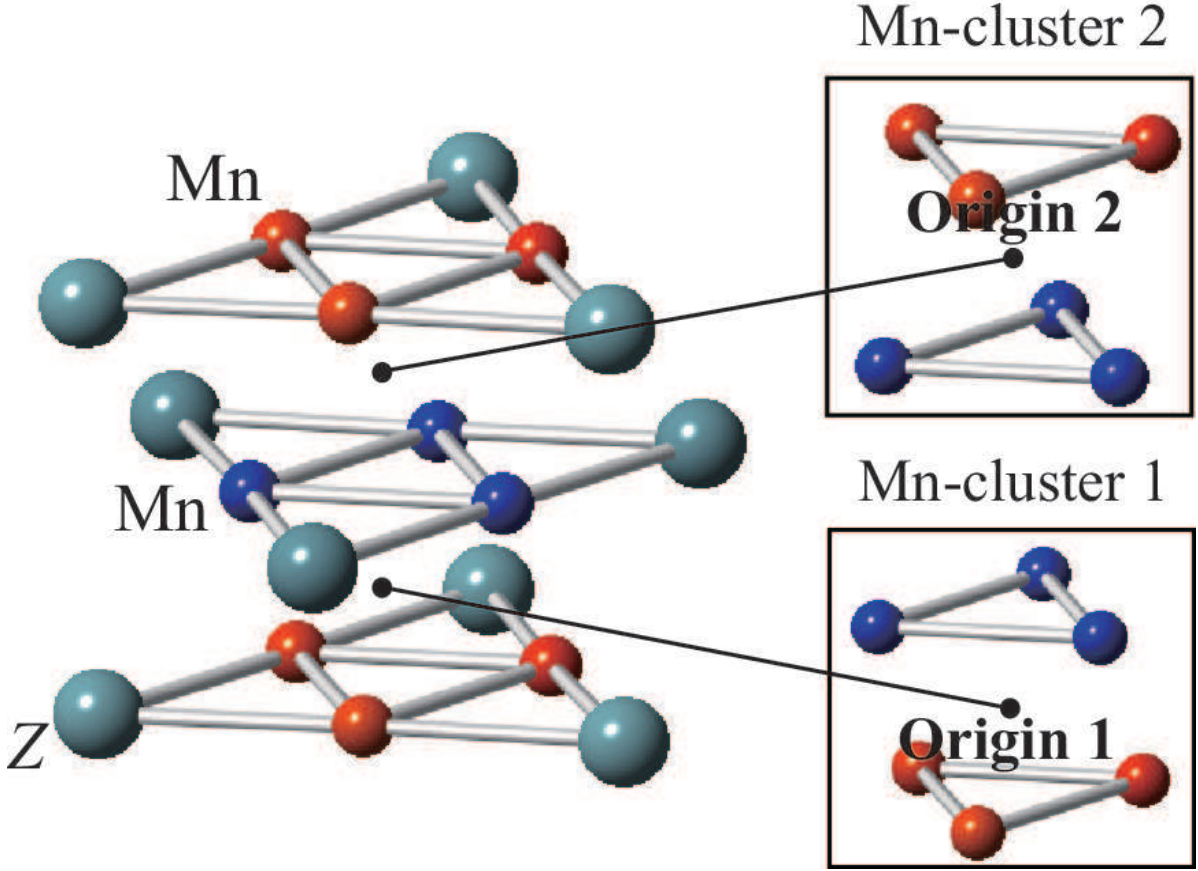}
        \caption{ (Color online) Crystal structure of the Mn$_3$$Z$ compounds and the
 inequivalent atomic clusters on which the cluster multipole moments are
 defined for the ${\bm q}={\bm 0}$ magnetic order (from Ref.\
 \cite{mtsuzuki2017}).}
\label{Fig:Cryst_Mn3Z}
\end{figure}


\subsection{Anomalous Hall effect and cluster multipole moments}

\begin{figure}[t]
 	\includegraphics[width=1.0\linewidth]{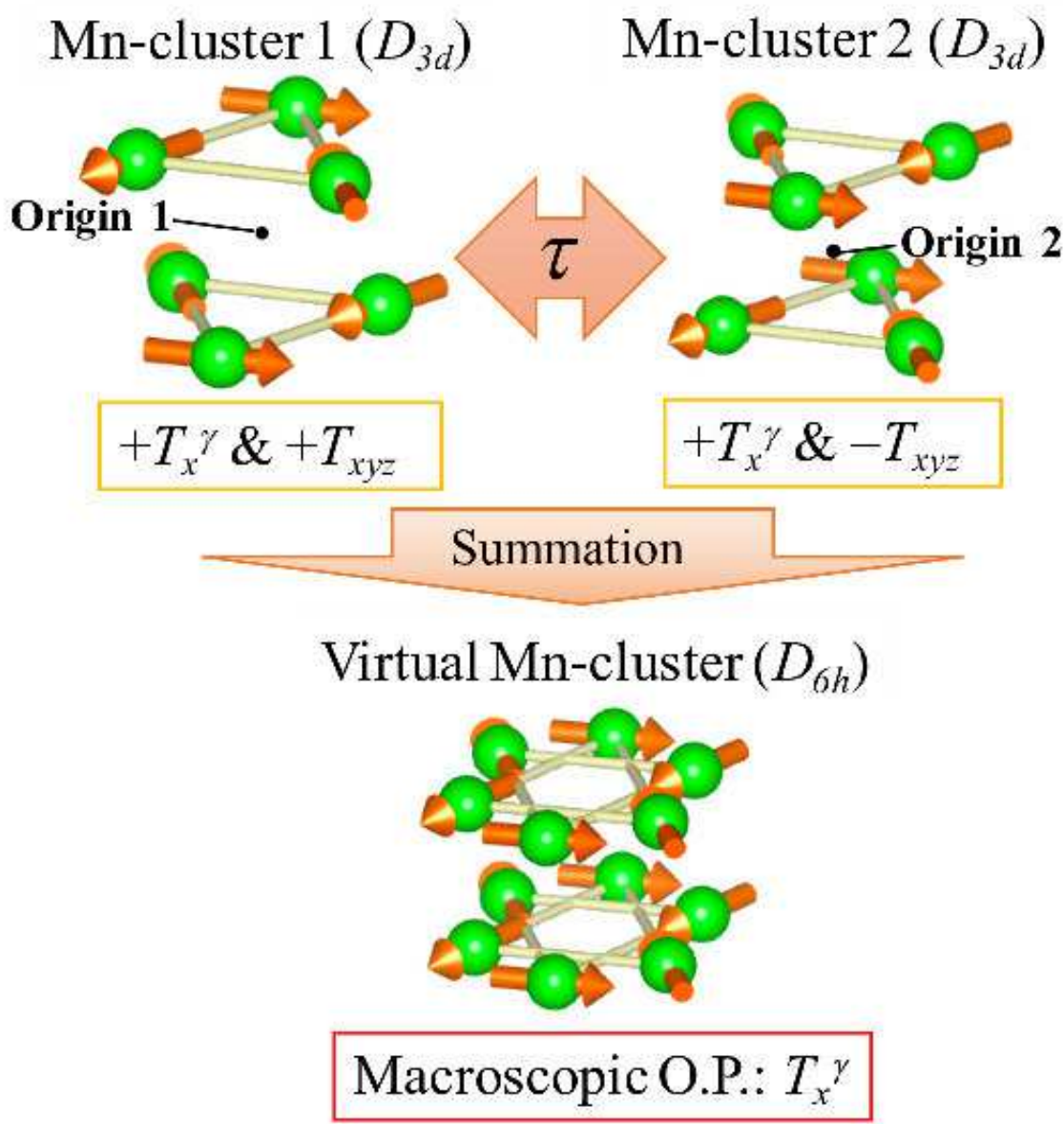}
        \caption{ (Color online) Schematics of the procedure to obtain the
 contribution of cluster multipole moments to the macroscopic order parameter (O.P.) in Mn$_3$$Z$.}
\label{Fig:macroCMP_Mn3Z}
\end{figure}

Large anomalous Hall conductivities (AHC) have recently been observed  in Mn$_3$$Z$ ($Z$=Sn, Ge)
  that are noncollinear antiferromagnets with vanishingly small magnetization and, according to conventional
  understanding, should exhibit negligible anomalous Hall
  conductivities~\cite{nakatsuji2015,kiyohara2016,nayak2016}.
   Mn$_3$$Z$ crystallizes into the hexagonal structure that
  belongs to the space group $P6_3/mmc$ (No.\ 194, Fig.\ \ref{Fig:Cryst_Mn3Z}).
 The nonsymmorphic space group $P6_3/mmc$ is decomposed in the symmorphic space
  group $P\bar{3}m1$ such as $P6_3/mmc=P\bar{3}m1+\{C_{2z}|{\bm
  \tau}\}P\bar{3}m1$.
  Following the procedure discussed in Sec.\ \ref{Sec:DefCMP}, Mn atomic
  clusters in the Mn$_3$$Z$ crystal are classified into two types, which
  consist of the atoms related to each other by the point group
  symmetry $D_{3d}$ ($\bar{3}m1$), with the representative origins shown
  in Fig.\ \ref{Fig:Cryst_Mn3Z}. The cluster
  multipole moments are calculated on each cluster using Eq.\
  (\ref{Eq:CMP}), leading to the finite cluster octupole moments
  $T_{x}^{\gamma}$ and $T_{xyz}$ as the lowest rank cluster multipole
  moment, and the macroscopic contribution is obtained by summing
  up the cluster multipole moments of these clusters following
  Eq.\ (\ref{Eq:CMPsum}).
  Figure \ref{Fig:macroCMP_Mn3Z} shows an image of the procedure to determine
  the macroscopic order parameter of the cluster octupole moment for the
  noncollinear antiferromagnetic structure of Mn$_3$$Z$.
 The procedure of summing up the cluster multipole moments of
  the two types of atomic clusters corresponds to neglect the
  translation degree of freedom which relates the different clusters
  (see Fig.\ \ref{Fig:macroCMP_Mn3Z}), and the resultant cluster
  multipole moments reflect the macroscopic symmetry breaking, i.e. the
  breaking of the paramagnetic $D_{6h}$ point group symmetry, due to the presence of  the magnetic configuration in the crystal.

 The noncollinear antiferromagnetic structure of hexagonal
 Mn$_3$Sn has thus been found to be
  characterized by the rank-3 cluster octupole moment which notably, belongs to the same
  irreducible representation as the rank-1 cluster dipole moment that 
  is obtained for the ordinary collinear ferromagnetic alignment of
  the magnetic moments~\cite{mtsuzuki2017}.
  Hence, these collinear ferromagnetic and noncollinear antiferromagnetic orders in fact  break the same
  magnetic symmetry and belong completely to the same magnetic space
  group. This is the reason, from the view point of symmetry, why the noncollinear
  antiferromagnetic order of Mn$_3$$Z$ can induce an AHC with the vanishingly small dipole magnetizations.
  On the other hand, the parasitic dipole magnetization can always
  appear for the AFM configurations that induce the AHC, as observed
  experimentally for Mn$_3$$Z$~\cite{tomiyoshi1982,brown1990,yamada1988,nayak2016,kiyohara2016}, since the
  appearing symmetry
  conditions both for the dipole magnetization and for the AHC are totally the same when the S-O coupling is considered~\cite{mtsuzuki2017}.
 Importantly, the magnetic symmetry breaking is dominantly caused by
 the antiferromagnetic configuration that is classified by the cluster
  octupole moments, and the small dipolar magnetization is not required to
  induce the large AHC observed in Mn$_3$$Z$.

\begin{figure}[t]
\begin{center}
	\includegraphics[width=0.6\linewidth]{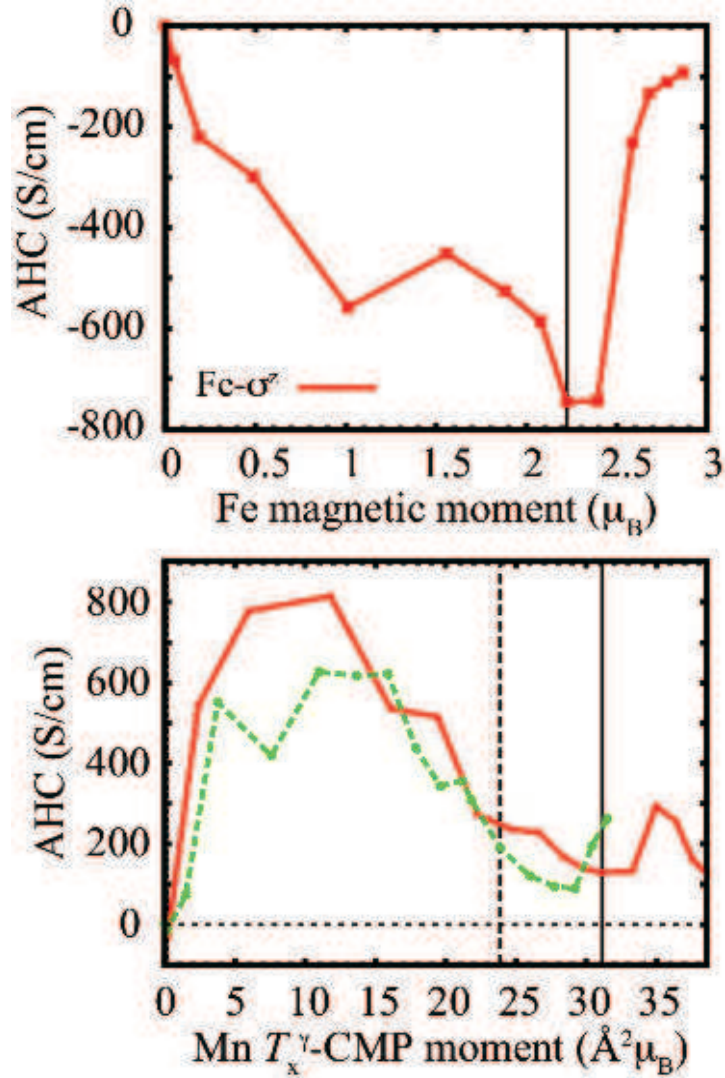}
	\caption{ (Color online) Calculated dependence of the anomalous Hall
 conductivity (AHC) on the dipole magnetic moment of BCC Fe (top panel)
	and on the cluster magnetic octupole moment of noncollinear
 antiferromagnetic Mn$_3$$Z$ (Z=Sn, Ge) (bottom panel). (from
 Ref.~\cite{mtsuzuki2017})
  The moments obtained by the
 first-principles calculations are shown by the solid lines for BCC Fe
 and Mn$_3$Sn and by the dashed line for Mn$_3$Ge.
}
	\label{Fig12}
\end{center}
\end{figure}
   The meaningfulness of the cluster multipole expansion can be exemplified 
   through direct \textit{ab initio} calculations of the AHC in relation to the multipole moments.
   The AHC can be calculated from a ${\bm k}$-space integration of the Berry
 curvature for the energy bands below the Fermi energy:
\begin{eqnarray}
\sigma_{\alpha\beta}=-\frac{e^2}{\hbar}\int
\frac{d{\bm k}}{(2\pi)^3}\sum_{n}f(\varepsilon_n({\bm
k})-\mu)\Omega_{n,\alpha\beta}({\bm k})\ ,
\label{Eq:sigma}
\end{eqnarray}
  where $\sigma_{\alpha \beta}$ is the off-diagonal anti-symmetric conductivity, $n$ is the band index, $f$ the Fermi function, $\mu$ the chemical potential, 
   and $\alpha$, $\beta=x$, $y$, $z$ with
  $\alpha\ne\beta$.
  The Berry curvature for the AHC is defined as
\begin{eqnarray}
  \Omega_{n,\alpha\beta}({\bm k})=-2{\rm Im} \sum_{m\ne
 n}\frac{v_{\alpha,nm}({\bm k})v_{\beta,mn}({\bm
 k})}{[\varepsilon_{m}({\bm k})-\varepsilon_{n}({\bm k})]^2},
\label{Eq:omega}
\end{eqnarray}
 from the Kubo linear-response formula~\cite{thouless1982, fang2003,
 wang2006}, where $v_{\alpha,mn}$ are the velocity matrix elements and $\varepsilon_n ({\bm k})$
 the band energies. These can conveniently be obtained from a
  downfolding of \textit{ab initio} calculated bands to a tight binding
 model that, discussed in Sec.\ \ref{Sec:tightbinding}, is useful to
 investigate the topological property of the electronic structure, such
 as Berry curvature and monopole charge, since the obtained tight
 binding model has a continuous phase factor of the eigenfunctions in
 ${\bm k}$-space.

    Figure \ref{Fig12} shows the \textit{ab initio} calculated dependence of the AHC on the size of the dipole magnetic moment in BCC Fe 
 and the size of the cluster octupole moment of AFM Mn$_3$$Z$~\cite{mtsuzuki2017}.
 The figure clearly establishes that the AHC depends on the cluster
 octupole moment of AFM Mn$_3$$Z$ similar to the dependence
  of the AHC on the magnetic dipole moment of ferromagnetic Fe.


\section{Concluding remarks}

The multipole framework, which has already been employed for a long time in various branches of physics, has recently proven to be a very fruitful concept, too,
in condensed matter physics. In the last three decades it has already become established that unconventional phases can be brought about by long-range ordering of multipole moments that have a higher rank than the conventional magnetic dipoles. A striking such example is the magnetic multipolar ordered phase of NpO$_2$, the origin of which could only be uncovered after more than three decades of intensive investigations \cite{santini2009}. 
Although the number of materials exhibiting long-range multipolar ordered phases is presently relatively small, these materials are precisely the ones that permit mankind to study the baffling interplay of several exotic order parameters and unveil as yet unknown aspects of fundamental interactions.

From the perspective of the electronic structure theory of solids it is gratifying that modern \textit{ab initio} calculation techniques are now accurate enough and capable to predict from first-principles multipolar ordered states in solid state materials \cite{mtsuzuki2010,mtsuzuki2013}.
These first-principles calculations are a powerful tool to provide deeper insight in the formation of multipole order in real materials. While CEF theory can often already explain the single-ion properties, the hybridizations between orbitals of neighboring atoms that occur in materials are not accounted for. However, these are important as precisely these lead to the formation of long-range order of multipoles in crystals. In the same way as long-range dipole magnetic order requires Heisenberg-Dirac exchange or Anderson super-exchange interactions to become stabilized at elevated temperatures, so do higher-order multipoles require a multipole-multipole exchange interaction for their stabilization \cite{mtsuzuki2010}. These exchange interactions can now be calculated using state-of-the-art electronic structure methods. 
It is however evident that there is still much uncharted territory in the \textit{ab initio} theory of multipole order. For comparison, about two decades ago it became feasible to compute \textit{ab initio} the Heisenberg exchange constants of elemental ferromagnets as Fe, Co, and Ni, and by mapping onto an effective  Heisenberg dipole-moment Hamiltonian, compute the Curie ordering temperatures.  Also the temperature dependence of the dipole order parameter could be explained (see e.g.\ Ref.\ {\cite{uhl1996,halilov1997}}).
This comparison exemplifies that there is still much unexplored physics; in particular, the step to compute multipolar ordering temperatures and the temperature-dependence of the order parameter is yet to be made.

A second conclusion derived from the recent research on magnetic multipoles is that there is a huge potential for novel, exotic phenomena. One such example is superconductivity mediated by multipole fluctuations. The possibility of superconductivity driven by spin fluctuations has been discussed much in relation to heavy-Fermion superconductors and high-transition temperature cuprates \cite{sato2001,moriya2000}.
Even though it is difficult to demonstrate superconductivity caused by
spin fluctuations, a new horizon opening up would be the possibility of
superconductivity mediated by multipole fluctuations. In this respect,
the origin of the unconventional superconductivity in URu$_2$Si$_2$ that
emerges only out of the HO phase is not understood yet. And as discussed
in this survey, magnetic multipole fluctuations could play a role for
the superconductivity in the archetypal heavy-Fermion superconductor
CeCu$_2$Si$_2$\cite{ikeda2015}.

A third, more general conclusion that can be drawn from recent
developments in cluster multipole theory, is that the multipole
description has important bearings for the understanding of macroscopic
phenomena in magnetic crystals. The new emerging understanding is that
the multipole moments pertaining to the magnetic structure characterize
magneto-electric phenomena. Thus, the cluster multipole theory offer
fresh insight in how unexpected macroscopic properties can become
realized \cite{mtsuzuki2017}.
These recent developments illustrate the power of the cluster multipole
expansion of the magnetic structures in crystal by combining symmetry arguments. This then in turn opens up for 
as yet unexplored ways of predicting materials with unusual properties,
based purely on the analysis of their crystal and magnetic point group
symmetries.
Even more, beyond the possibility of performing such analyses, there are
prospects to search for such materials by employing numerical search
routines in the future. There exist already extensive data bases of
crystal structures that need now to be combined with the allowed
magnetic configurations to provide multipole classifications. In
combination with machine learning this might initiate new pathways for
the discovery, and even design, of novel materials with intricate
properties in the coming years.


\section*{Acknowledgment}
We thank N.\ Magnani, T.\ Takimoto, Y.\ Matsuda, T.\ Shibauchi, T.\
  Koretsune, M.\ Ochi, R.\ Arita for the collaborative works and R.\
  Caciuffo, G.\ H.\ Lander, J.\ A.\ Mydosh, H.\ Kusunose, S.\ Nakatsuji,
  and H. Harima for fruitful discussions.
 This work has been supported by the JSPS KAKENHI (Grant Numbers
  JP15K17713, JP15H05883 (J-Physics), and JP16H04021, JP16H01081,
  JP15H05745, JP15H02014, JP24540369), the Swedish Research Council (VR), the K.\ and A. Wallenberg Foundation (Grant No.\ 2015.0060), and the Swedish National Infrastructure for Computing (SNIC).\\


\section*{Appendix: Symmetry of multipolar ordered states}
\label{Sec:SymMPO}
Symmetry analysis is a crucial step to understand the electronic states
in complex order phases. Indeed, the experimental identification of
 an ordered state is often equivalent to detect the physical
object or scattering  from it accompanied by a conspicuous symmetry breaking.
   Furthermore, setting an appropriate constraint on the magnetic symmetry is crucial to obtain a
   multipolar ordered state of interest as a convergent solution in selfconsistent electronic structure calculations.
   As discussed in Sec.\ \ref{Sec:Method}, an LDA+$U$
 calculation for a multipolar ordered state requires the introduction of an
 appropriate symmetry breaking that permits the presence of the multipolar order.

  The symmetry operations which preserve the multipolar order are
 determined from the transformation properties of the local multipole
 moments, classified according to the irreducible representations of the
 point group, and its configuration on the atoms which electron and spin density induce the
 multipole moments in the crystal. The symmetry property of the multipolar
 order can be characterized by the magnetic space group (MSG, or
 Shubnikov group) as well as that of ordinary magnetic order.
  Once we identify the MSG of the multipolar order,
  we can further obtain information on local site symmetry of each
 atomic site in crystal, leading to conclusions about
 possible atomic distortion or secondary order parameters induced with
 the primary multipole order~\cite{mtsuzuki2014}. Since the magnetic
 multipole moment whose rank is higher than one is often difficult to
 directly detect experimentally, information on the secondary order
 induced with the primary magnetic multipole order is often crucial to
 identify the ordered states.

  Although there are exhaustive numbers of MSGs, these are all
 classified according to the following four types:~\cite{bradley1972_book}
\begin{eqnarray}
 \mathcal{M} &=&\mathcal{G}\ \ \ \ \ \ \ \ \ \ \ \ \ \ \ \ \ \ \ \ \ \ \ \ \ \ \ {\rm Type~ I,}
\label{eq:Fedrov}\\
 \mathcal{M} &=&\mathcal{G}+\theta \mathcal{G}\ \ \ \ \ \ \ \ \ \ \ \ \ \ \ \ \ \ {\rm Type~ II,}
\label{eq:TypeII} \\
 \mathcal{M} &=&\mathcal{H}+\theta(\mathcal{G}-\mathcal{H})\ \ \ \ \ \ {\rm Type~ III,}
\label{eq:TypeIII}\\
 \mathcal{M} &=&\mathcal{G}+\theta\{E\mid{\bm \tau}\}\mathcal{G}\ \ \ \ \ \ \ \ {\rm Type~ IV.}
\label{eq:TypeIV}
\end{eqnarray}
 Here $\mathcal{M}$ is the MSG, $\theta$ is the time-reversal
operation, $\mathcal{G}$ is the ordinary space group, which do not have
the symmetry operations including the time-reversal operation,
$\mathcal{H}$ is a subgroup of the ordinary space group
$\mathcal{G}$ whose number of
symmetry operations is half of that of 
$\mathcal{G}$ and $\mathcal{G}-\mathcal{H}$ contains no pure
translations, $\{R\mid{\bm \tau}\}$ is a successive transformation of
a point group operation $R$, which is the identity operator $E$ in Eq.\
(\ref{eq:TypeIV}), and a translation ${\bm \tau}$.
The terms related to the time-reversal operation in Eqs.\
(\ref{eq:TypeII}), (\ref{eq:TypeIII}), and (\ref{eq:TypeIV}) are called
the anti-unitary part of the MSG.
  The type II MSG is called grey space groups, and the nonmagnetic
  orders, including pure electric multipole orders, belongs to the grey
  groups since these orders preserve the pure time-reversal symmetry.
General magnetic multipolar ordered states that break pure time-reversal 
symmetry, belong either to type I, III, or IV MSG.
 As discussed below, the difference of the MSG's types of magnetic
 multipolar ordered states leads to 
 qualitatively different consequences for the secondarily induced
 electric order parameters.

The type I MSG is called Fedorov group. 
 In magnetic ordered states which belong to the Fedorov group $\mathcal{M}$
 ($=\mathcal{G}$), the symmetry of the charge distribution is characterized
 by the space group $\mathcal{G}$ and is not distinguished from that of
 the magnetic order. 
The 3${\bm q}$ antiferroic multipolar ordered phases of UO$_2$
and NpO$_2$ belong to the MSGs $Pa\bar{3}$ and $Pn\bar{3}m$,
respectively~\cite{nikolaev2003}, which both belong to the Fedorov group in Eq.\ (\ref{eq:Fedrov}).
  The transverse 3${\bm q}$ magnetic $T_{1g}^{-}$ dipole order
of UO$_2$ and the longitudinal 3${\bm q}$ magnetic
$T_{2g}^{-}$ multipole order of NpO$_2$ induce a 3${\bm q}$ ordering of the electric
$T_{2g}^{+}$ quadrupole moments ($Q_{yz}$, $Q_{zx}$, $Q_{xy}$) as the
secondary order.
  Concerning the local symmetry, the $Pa\bar{3}$ MSG of UO$_2$
 leaves the oxygen atoms equivalent but allows for displacements of the
 oxygen sublattice since the Wyckoff position has one parameter,
 corresponding to the 8c site in $Pa\bar{3}$ space group. On the
 other hand, the $Pn\bar{3}m$ MSG of NpO$_2$ does not allow oxygen
 displacements but splits the oxygen atoms into two inequivalent atomic
 sites, corresponding to 2a and 6d sites of $Pn\bar{3}m$ space group~\cite{nikolaev2003}. 
This crucial difference between UO$_2$ and NpO$_2$ is confirmed by
NMR experiments~\cite{tokunaga2005b,tokunaga2006b}. 

   The type III and IV MSGs are called black and white space groups, respectively.
  The magnetic ordered states which belong to these MSGs clearly distinguish the symmetry of the charge
  distribution from that of the magnetic order through the time-reversal
  related symmetry operations of the anti-unitary part in Eqs.\
 (\ref{eq:TypeIII}) and (\ref{eq:TypeIV}), i.e. the symmetry of charge
 distribution belongs to $\mathcal{G}$ and by $\mathcal{G}+\{E\mid {\bm \tau}\}$$\mathcal{G}$ in the magnetic
  order of type III and type IV MSG, respectively, since the charge
  density is time-reversal even. 
    Remarkably, in the magnetic multipolar ordered states of type IV MSG, 
  the charge distribution is invariant under the pure translation operation
  ${\bm \tau}$, and the translation symmetry is broken only by the
  magnetic degree of freedom.
  The general consequence of the MSGs, discussed above, leads to a crucial difference
  between the low temperature magnetic multipolar phase of NpO$_2$ and
  the possible magnetic multipolar ordered states of URu$_2$Si$_2$ as the main
  candidates of the HO state.

  The space group analysis of the HO phase in URu$_2$Si$_2$ identified
 possible ${\bm q}=(0,0,1)$ antiferroic multipolar orders that preserve
 the symmetry which is difficult to detect experimentally
 \cite{harima2010,khalyavin2014,mtsuzuki2014}.
 In particular, the antiferroic orders of the magnetic multipole
moments which belong to one-dimensional IREPs ($A_{1g}^{-}$, $A_{2g}^-$,
 $B_{1g}^{-}$, $B_{2g}^{-}$) of the $D_{4h}$ point
group never break the crystal symmetry of the charge
distribution, which is a consequence of the fact 
that the magnetic multipolar states belong to type IV
 MSG~\cite{khalyavin2014,mtsuzuki2014}.
  Furthermore, since the magnetic dipole moments belongs to the
 $A_{2g}^-$ IREP, the $A_{1g}^{-}$, $B_{1g}^{-}$, and $B_{2g}^{-}$
 multipole moments contain no rank-1 magnetic dipole moment, and
 AF-$A_{1g}^{-}$ and AF-$B_{1g}^{-}$ orders especially do not induce the magnetic
 dipole moment at any atomic site in URu$_2$Si$_2$ from the symmetries~\cite{mtsuzuki2014}.
  Meanwhile, the $E_{g}^{-}$ multipole moments contain the in-plane rank-1 magnetic
 dipole moments, and AF-$E_{g}^{-}$ multipolar order can induce a charge deformation
 corresponding to the {\it ferroic} electric $E_{g}^{+}$ multipolar order
 and the magnetic dipole moments at all of the atomic site in
 URu$_2$Si$_2$~\cite{mtsuzuki2014}.
 The symmetry breaking of the charge distribution can be related to the
 experimentally observed nematic feature in the HO
 phase~\cite{okazaki2011,tonegawa2012,kambe2013,tonegawa2014,riggs2015},
 although the existence of the nematic feature is still controversial~\cite{saitoh2005,mito2013}.
  On the other hand, the induced  magnetic moments have not been
 experimentally confirmed as an intrinsic effect~\cite{takagi2012,das2013,metoki2013,ross2014}.

The charge distribution of AF-{\it magnetic} multipolar ordered states in
 URu$_2$Si$_2$ are thus clearly distinguished from the magnetic distribution
 in terms of symmetry, and the {\it magnetic} AF-$\Gamma^{-}$ multipolar ordered states
 always preserve the symmetry of the charge distribution higher than
 that of the corresponding electric AF-$\Gamma^{+}$ multipolar ordered states.
  From this perspective, the occurrence of the AF-{\it magnetic}
 multipolar ordered states is more effective to conceal the ordered states by
 preserving the charge symmetries higher than the ones of the electric
 multipolar ordered state.
This crucial difference  in the symmetry
property of the magnetic multipolar order could be related to the
difficulty of experimental detection of the symmetry breaking of the HO
 phase in URu$_2$Si$_2$.


\end{document}